\documentclass[12pt,draftclsnofoot,onecolumn]{IEEEtran}
\usepackage{amsmath}
\usepackage{amsfonts}
\usepackage{amssymb}
\usepackage{cite}
\usepackage{graphicx}
\usepackage{mathtools}
\usepackage{cases}
\usepackage{color}
\usepackage{bbm}

\usepackage{datetime}

\newtheorem{theorem}{Theorem}

\newtheorem{corollary}{Corollary}

\newtheorem{definition}{Definition}
\newtheorem{example}{Example}

\newtheorem{lemma}{Lemma}

\newtheorem{proposition}{Proposition}
\newtheorem{remark}{Remark}

\DeclareMathOperator*{\argmax}{arg\,max}
\DeclareMathOperator\erf{erf}

\DeclareMathOperator{\cA}{\mathcal{A}}

\DeclareMathOperator{\cL}{\mathcal{L}}

\DeclareMathOperator{\SIR}{\textrm{SIR}}
\DeclareMathOperator{\bR}{\mathbb{R}}
\DeclareMathOperator{\bP}{\mathbf{P}}\DeclareMathOperator{\bE}{\mathbf{E}}
\DeclareMathOperator{\bZ}{\mathbb{Z}}
\newcommand*\diff{\mathop{}\!\mathrm{d}}

\newcommand*\nnb{\nonumber}

\author{Chang-Sik Choi, Jae Oh Woo, and Jeffrey G. Andrews \IEEEcompsocitemizethanks{\IEEEcompsocthanksitem Chang-sik Choi, Jae Oh Woo, and Jeffrey G. Andrews are with the Wireless and Networking and Communication Group, Department of Electrical and Computer Engineering, The University of Texas at Austin, Texas, 78701, USA. (e-mail: chang-sik.choi@utexas.edu, jaeoh.woo@gmail.com, jandrews@ece.utexas.edu) 
}
\thanks{Part of this work has been presented in IEEE ISIT 2017 \cite{8006571} }}
\title{An Analytical Framework for Modeling a Spatially Repulsive Cellular Network}
\begin{document}
	\maketitle
\begin{abstract}
We propose a new cellular network model that captures both deterministic and random aspects of base station deployments.  Namely, the base station locations are modeled as the superposition of two independent stationary point processes: a random shifted grid with intensity $ \lambda_g $ and a Poisson point process (PPP) with intensity $ \lambda_p $.  Grid and PPP deployments are special cases with $\lambda_p \to 0$ and $\lambda_g \to 0$, with actual deployments in between these two extremes, as we demonstrate with deployment data.  Assuming that each user is associated with the base station that provides the strongest average received signal power, we obtain the probability that a typical user is associated with either a grid or PPP base station. Assuming Rayleigh fading channels, we derive the expression for the coverage probability of the typical user, resulting in the following observations. First, the association and the coverage probability of the typical user are fully characterized as functions of intensity ratio $ \rho_\lambda=\lambda_p/\lambda_g$.  Second, the user association is biased towards the base stations located on a grid. Finally, the proposed model predicts the coverage probability of the actual deployment with great accuracy.
	
\end{abstract}
\section{Introduction}
\subsection{Motivation and Related Work}
\par  The topological distribution of base stations (BSs) is a first order effect in terms of the signal-to-interference-and-noise ratio (SINR) of a cellular network.  BSs are usually intentionally spread out from one another, which is referred to as \emph{repulsion} in studies  \cite{lee2013stochastic,7110502,taylor2012pairwise,nakata2014spatial}.  However, those works did not analytically derive SINR expressions, primarily because it is quite difficult to do so for repulsive network model.  Two examples of actual BS deployments are given in Figures \ref{Data1} and \ref{Data2}, in which repulsion can be observed both subjectively and objectively, by using statistical metrics.  The goal of this paper is to propose a novel model and  analytical framework for cellular networks in view of repulsive BS deployments.

\subsection{Background: From Poisson to More General Models}
\par The modeling and analysis of BS locations using the Poisson point process (PPP) has become popular in the last five years because the independence of the PPP helps allow computable---and in some cases quite simple---SINR distribution expressions. The PPP-based models describe randomly distributed BSs and often yield a simple expression for the interference \cite{haenggi2009interference,baccelli2009stochastic}. In \cite{andrews2011tractable} the coverage probability of the typical user---meaning the probability that the SINR is above a certain value---was derived in a closed form, and was extended in \cite{Dhillon2012modeling,Jo2012heterogeneous,6171998} to multi-tier heterogeneous cellular networks (HCNs). The benefits of HCNs were investigated further in \cite{6463498} and \cite{ye2013user}. The PPP model was also used for finding the average rate \cite{di2013average}, obtaining an energy-efficient algorithm \cite{soh2013energy} and \cite{huang2014enabling}, and deriving the signal-to-interference (SIR) of millimeter cellular networks \cite{bai2015coverage}. A PPP approximation was studied in \cite{heath2013modeling,blaszczyszyn2013using,guo2015asymptotic,wei2016approximate}. However, the PPP model ignores the repulsive nature of spatial topology observed in cellular networks. 

\par In order to mathematically represent repulsion between BSs,  \cite{6841639,li2015statistical,ibrahim2013coverage} considered repulsive point processes given in \cite[Chapter 5]{chiu2013stochastic} and attempted to derive the SINR expressions analytically.  Several specific point processes were used. For example, \cite{6841639} considered the Ginibre point process, \cite{li2015statistical} provided tractable but complicated analysis for the more general determinantal point process, and \cite{ibrahim2013coverage} used the Mat{\'e}rn hard-core point process, which introduces an exclusion region around each BS.

\subsection{Contributions}

\begin{figure}
	\centering
	\includegraphics[width=0.8\linewidth]{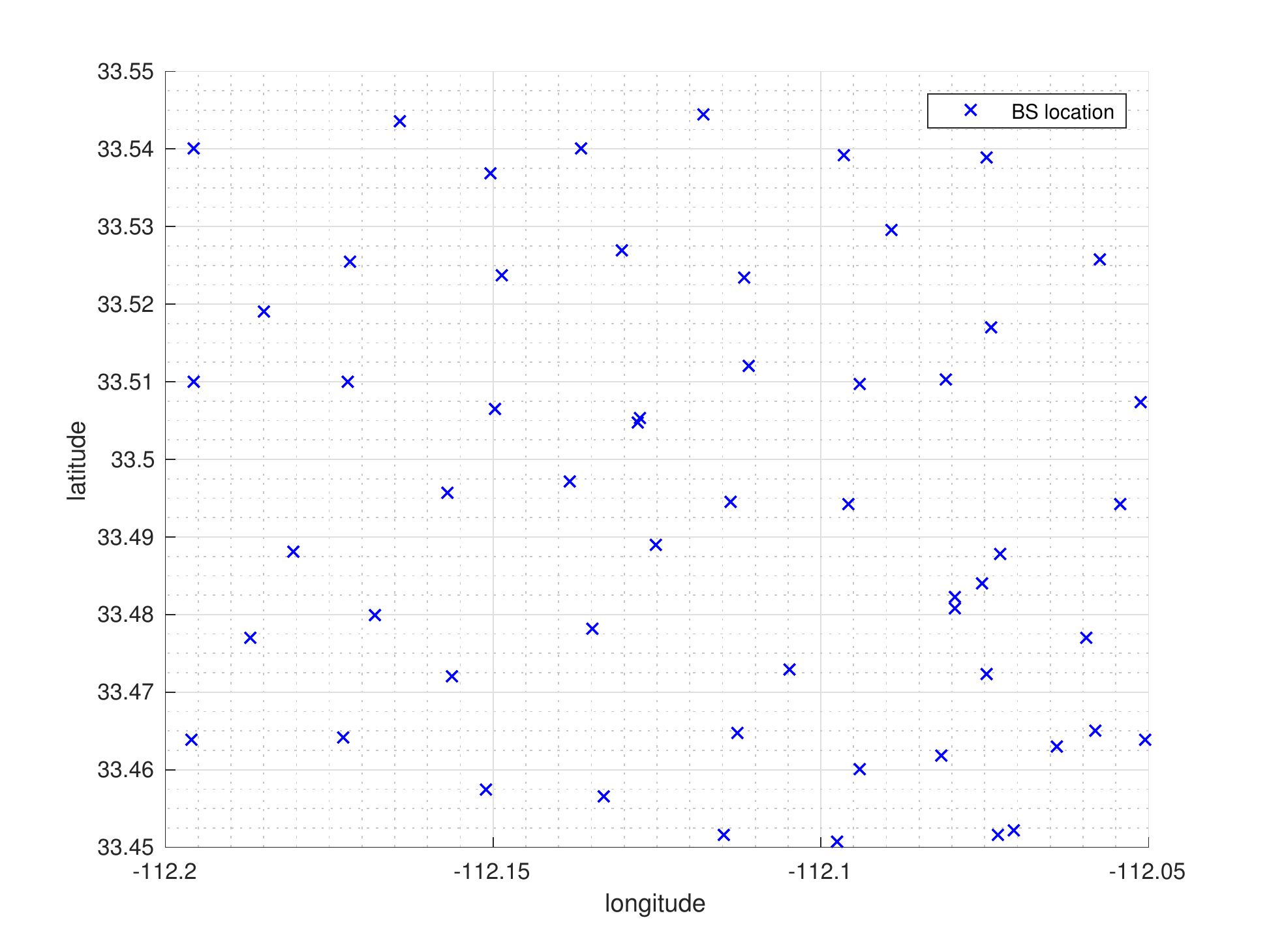}
	\caption{Actual BS deployment for the area of latitude and longitude $ [33.45,33.55]\times [-112.2,-112.05] $. The pair correlation value is 0.90 and the corresponding estimated intensity ratio is 2.3.\label{Data1}}
\end{figure}

\begin{figure}
	\centering
	\includegraphics[width=0.9\linewidth]{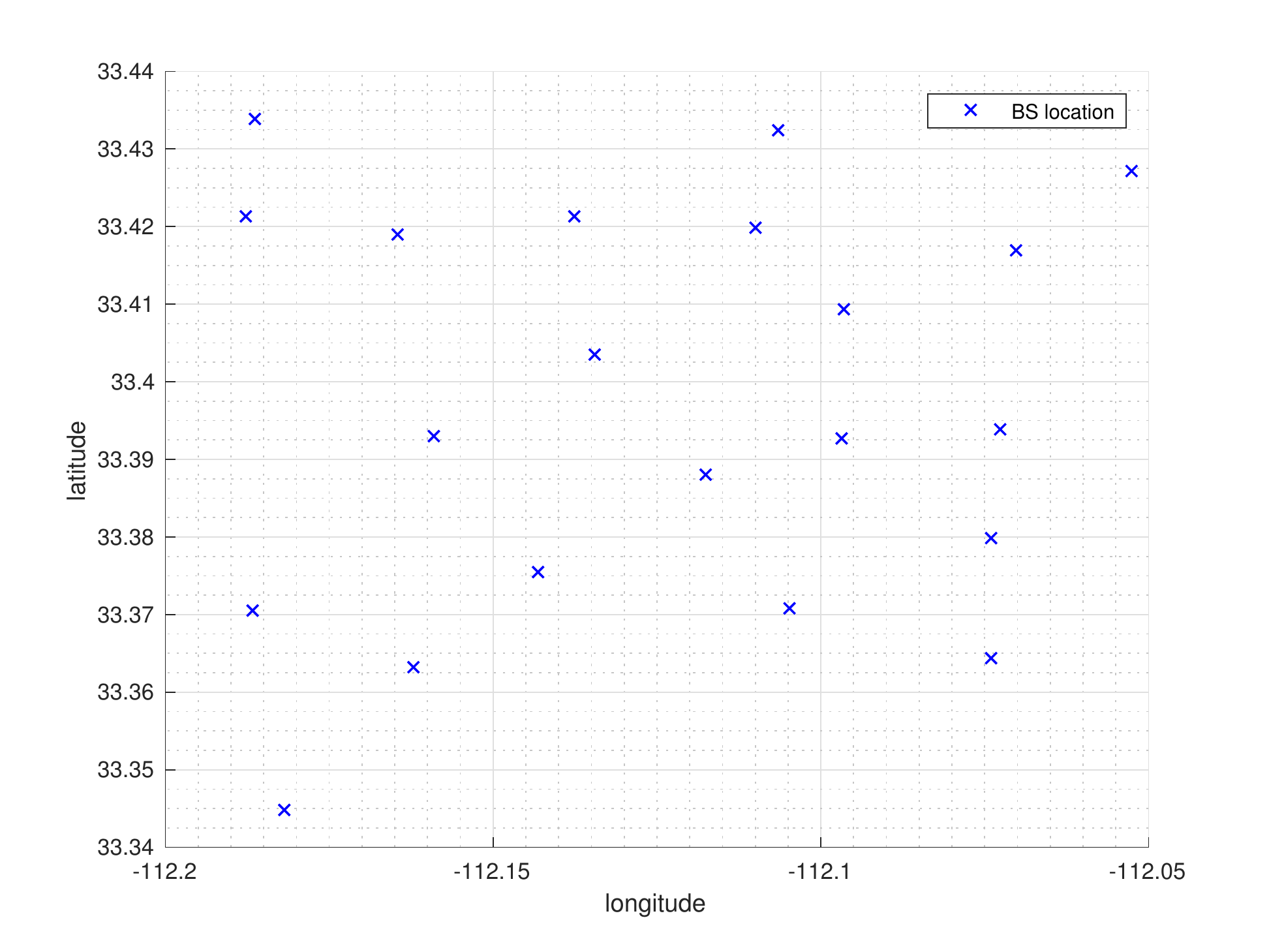}
	\caption{Actual BS deployment in the area of latitude and longitude $ [33.34,33.44]\times[-112.2,-112.05] $. The pair correlation value is 0.776 and the corresponding estimated intensity ratio is 1.11}\label{Data2}
\end{figure}

\begin{figure}
	\centering
	\includegraphics[width=1\linewidth]{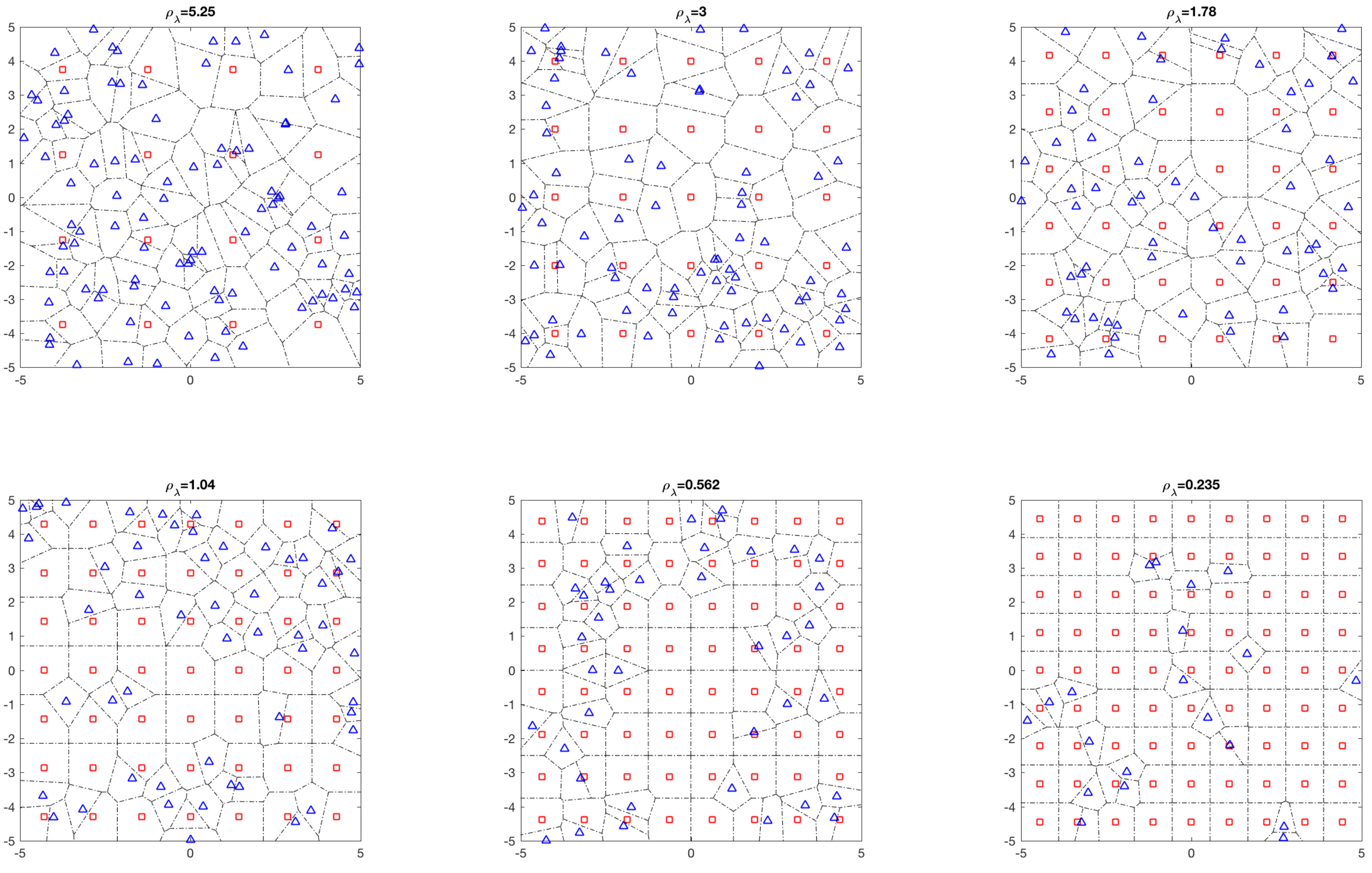}
	\caption{The grid BSs and PPP BSs are demonstrated by squares and triangles, respectively. The interpolation of the proposed model is described for multiple values of $ \rho_\lambda$. We maintain the total number of BSs to minimize the edge effect. The dashed lines indicate the Voronoi boundaries.}
	\label{fig:rho}
\end{figure}

This paper proposes to model BS locations using a stationary point process that is constructed by the superposition of two independent stationary point processes.  The first is a standard PPP with density $\lambda_p$ and the second is a random shifted grid with intensity $\lambda_g$.  The realizations of the proposed point process and its association regions are given in Figure \ref{fig:rho} for multiple values of $ \rho_\lambda $.  Possible real-world examples where such a model could be applicable including
\begin{itemize}
	\item A vehicular network where fixed infrastructure modeled by grid and randomly located vehicles modeled by Poisson points can transmit information to other vehicles.
	\item A device-to-device (D2D) network with fixed BSs and randomly scattered D2D devices; a given mobile receiver is able to receive transmissions from either one.
	\item A HCN with repulsively (nearly in a grid) macro BSs overlaid with small cells randomly scattered over space.
\end{itemize}
As can be seen heuristically in Figures \ref{Data1} and \ref{Data2}, actual data also conforms to such a model where BSs are placed both randomly and regularly.

%

The theoretical contributions of this paper are as follows.

\par \textbf{Analytical framework capturing repulsion between BSs.}  We propose to model cellular networks as a combination of two extreme sub-structures. The repulsive BSs of transmit power $ p_g $ are modeled by a random shifted grid with intensity $ \lambda_g $ and the random BSs with transmit power $ p_p $ are modeled by a PPP with intensity $ \lambda_p $. We capture repulsion by the intensity ratio $ \rho_\lambda = \frac{\lambda_p}{\lambda_g}$. In fact, any network model between the grid and the PPP can be reproduced by simply varying $ \rho_\lambda $. When $ \rho_\lambda \to 0 $ the proposed model describes the lattice BS model: meanwhile, when $ \rho_\lambda \to\infty $ the proposed model corresponds to the Poisson BS model. The interpolation is demonstrated in Figure \ref{fig:rho}. Consequently, the proposed model is readily applicable to practical deployments compared to models with only lattice \cite{haenggi2010interference} or PPP BSs \cite{andrews2011tractable}.
\par \textbf{Derivation of the nearest distance distribution, user association probability, interference, and the SIR coverage probability.}  First,  the nearest distance distribution of the proposed stationary point process is characterized. Then, by assuming a typical user is associated with the BS that provides the strongest average received power, the association probability of the typical user is derived. Assuming Rayleigh fading channels, the Laplace transform of interference seen at the typical user is derived and the expression for the coverage probability of the typical user is obtained. The association probability, the interference, and the coverage probability are described by functions of system parameters including the intensity ratio $ \rho_\lambda $ and the power ratio $ \eta =\frac{p_p}{p_g}$.  
\par \textbf{Insights and observations from the proposed model.}  First, in user association, a bias towards the repulsive BSs is observed.
For instance, when $ \rho_\lambda = \eta = 1$, the typical user at the origin is more likely to be associated with a grid BS than a randomly located BS. Second, the proposed model is robust and consistent for different values of $ \alpha.$ We numerically show that the proposed model predict the coverage probability of the actual deployment very accurately.



\par The rest of this paper is organized as follows. Section \ref{s:2} discusses the proposed system model and defines performance metrics. Section \ref{s:3} investigates mathematical properties of the proposed point process and its tessellation. Section \ref{s:4} discusses the user association, interference, the coverage probability of the typical user, and its bounds. Section \ref{s:7} concludes the paper.

\section{System Model}\label{s:2}
\begin{table}
	\caption{Key Network Parameters}
\begin{center}
	\begin{tabular}{|l|l|}\hline
		Parameter  &Notation \\ \hline\hline
		Random shifted grid BS (intensity) &$ \Phi_g (\lambda_g)$ \\
		Homogeneous Poisson BS (intensity)&$ \Phi_p (\lambda_p) $\\
		The proposed point process & $ \Phi=\Phi_g\cup \Phi_p $\\
		Intensity ratio & $ \rho_\lambda=\lambda_p/\lambda_g $\\
		Fading random variable & $ H $\\
		Transmit power ratio & $ \eta=p_p/p_g $\\
		Nearest distance from the origin to the $ \Phi_g $ & $ R_g $ \\
		Nearest distance from the origin to the $ \Phi_p $ & $ R_p $ \\
		The smallest element of grid & $ S_0\in\bR^2 $\\
		The area of a set $ A $ & $ \nu_2(A) $\\
		An uniform random variable on a set $ A $ & \text{Uniform}[$ A $]\\
		The area of the typical Voronoi cell w.r.t. grid (Poisson) BSs & $ \bar{\mathcal{V}_g} (\bar{\mathcal{V}_p}) $\\
		The event that typical user is associated with $ \Phi_g(\Phi_p) $ & $ \mathcal{A}_g (\mathcal{A}_p) $\\
		\hline
	\end{tabular}
\end{center}

\end{table}



\par 
\subsection{Spatial Model}
To model the locations of the grid BSs, a random shifted grid is used. The random shifted grid is given by shifting the vertexes of the standard square 2-d grid via a single uniform random variable while maintaining the entire grid structure. The locations of the BSs are given by
\begin{align}
\Phi_g=\sum_{k \in \mathbb{Z}^2}\delta_{{s\cdot k+U}}\label{eq:randomgrid},
\end{align}
where $ \delta_{x} $ denotes the dirac-delta function at $ x $, $ s $ denotes the distance between the nearest vertexes, and $ U $ denotes the uniform random variable on $ \left[-\frac{s}{2},\frac{s}{2}\right]^2 $. The above expression shows that all points of the square grid are shifted by \emph{a random variable} $ U $ as a group. This is different from the perturbed lattice produced by shifting each point using i.i.d. uniform $ U_i $. The perturbed lattice is inappropriate for the modeling purpose since the deterministic structure between BSs no longer exists and repulsion between BSs is diluted. On the other hand, the random shifted grid is a stationary point process because the joint distribution,
\begin{equation}
\{\Phi_g(\mathcal{B}_1+x),\Phi_g(\mathcal{B}_2+x),\ldots,\Phi_g(\mathcal{B}_k+x)\}, 
\end{equation}
does not depend on the location of $ x\in\bR^2 $
for any finite Borel set $ \mathcal{B}_k \in \mathcal{B}(\bR^2) $  \cite[Definition 3.2.1.]{daley2007introduction}.
Since it is a stationary point process, it admits the intensity parameter denoted by $ \lambda_g $. It is given by the average number of points in a unit area $ \lambda_g=  1/s^2 $. Throughout this paper, the random shifted grid is referred to as stationary grid or just \emph{grid}. On the same space that $ \Phi_g $ is considered, an independent PPP with intensity parameter $ \lambda_p $ is considered
\begin{equation}
\Phi_p=\sum_{i\in\mathbb{N}}\delta_{X_i}.
\end{equation}

\par Since the two point processes are stationary, the superposition of the point processes also yields a stationary point process. The proposed point process is given by   \begin{equation}
\Phi=\Phi_g\cup \Phi_p.  
\end{equation}
its intensity parameter is $\lambda=\lambda_g+\lambda_p$. We introduce a prefatory parameter $ \rho_\lambda $ defined by 
\begin{equation}\label{ratio}
\rho_\lambda=\frac{\lambda_p}{\lambda_g},
\end{equation}
which will be seen repeatedly in this paper. Note that repulsion between BSs can be modeled in many different ways as presented in \cite{nakata2014spatial,6841639,li2015statistical,ibrahim2013coverage}; this paper proposes one way of characterizing repulsion using \eqref{ratio}. Under the proposed framework, if $ \rho_\lambda $ tends to zero, the proposed network describes the grid almost everywhere. If $ \rho_\lambda $ tends to infinity, the proposed network describes the PPP almost everywhere. In this sense, the proposed model is designed to interpolate the grid and the PPP. It is clarified in Figure \ref{fig:rho}.

\par In addition to the proposed point process $ \Phi $, another stationary independent PPP $ \Phi_r $ is considered to describe the locations of downlink cellular users. Under the stationary framework, a typical user located at the origin is considered. 
\begin{remark}
	The proposed point process, in particular the random shifted grid, is able to address repulsion between pairs of BSs  by changing the intensity $ \lambda_g $ w.r.t. $ \lambda_p $. This compares favorably to the models using perturbed lattice \cite{6214312} where the repulsion is softly induced by having only one point per grid. In other words, the perturbed lattice model potentially places two BSs very close to each other, which is observed rarely in actual cellular deployments. In addition, the proposed model captures the repulsion by a single scalar parameter $ \lambda_g $. This approach contrasts to the Ginibre point process \cite{6841639} or determinantal point process \cite{li2015statistical} where the repulsions are modeled for all pairs of points and require a multi-dimensional parameter.
\end{remark}

\subsection{Transmission Model}
The received signal power at distance $R$ is given by  
$ {P H}R^{-\alpha}  $
where $ P $ denotes the transmit power,  $ H $  denotes Rayleigh fading that follows exponential distribution with mean one, and $ R^{-\alpha} $ denotes path loss with $ \alpha>2 $. The transmit power $ P $ of the BS at $ X $ is given by   
\[P_X=\begin{cases}
p_g. & \textrm{for $ X \in \Phi_g$}\\
p_p. & \textrm{for $ X \in \Phi_p$}
\end{cases}\]
This paper assumes that cellular downlink users are associated with BSs that provide the strongest average power. The BS associated with the typical user is given by
\begin{align}
X^\star &=  	\argmax_{X_i\in\Phi} \bE\left[\frac{P_{X_i}H_{X_i}}{\|X_i-0\|^{\alpha}}\right]=\argmax_{X_i\in\Phi} \frac{P_{X_i}}{\|X_i\|^{\alpha}},\label{eq:xstar}
\end{align}
where the subscripts on $ H $ and $ P $ are used to specify the transmitter and $ \|\cdot\| $ denotes the Euclidean distance. 
Using $ X^\star $, the received signal power of the typical user at the origin is given by
\begin{align}
S={P_{X^{\star}} H_{X^\star} }{\|X^\star\|^{-\alpha}},
\end{align}
and the interference is given by
\begin{align}
I=\sum_{X_i\in\Phi \setminus \{X^\star \} } {P_{X_i}H_{X_i}}{ \|X_i\|^{-\alpha}}\label{eq:System-I}.
\end{align}

\subsection{System Performance Metric}
In this paper, we use the coverage probability of the typical user to assess the coverage probability of all users averaged across the space. Provided that noise power at the user is minuscule compared to the received signal power or the interference powers, which is commonly assumed in practical dense cellular networks, the coverage probability of the typical user is captured by the SIR at the typical user 
\begin{equation}
\SIR=\frac{S}{I}=\frac{{P_{X^{\star}} H }{\|X^\star\|^{-\alpha}}}{\sum_{X_i\in\Phi \setminus \{X^\star \} } {P_{X_i}H}{ \|X_i\|^{-\alpha}}}\label{eq:SIR},
\end{equation}
where subscripts on $ H $ are omitted.
The typical transmission at the typical user is successful if the above typical SIR is greater than a threshold $ T $. The coverage probability of the typical user is written as 
$  p_{\textrm{cov}}=\mathbf{P}^0( \SIR\geq  T)\label{eq:cover} $ where we use the Palm notation.
\section{Properties of the Proposed Model: Nearest Distance Distribution and the Voronoi Tessellation}\label{s:3}
This section discusses mathematical properties of the proposed point process for BSs using stochastic geometry. 
\subsection{Nearest Distance Distribution}The distances from the typical point to the nearest points of $ \Phi_g $ or $ \Phi_p $ are defined by 
\begin{align}
R_g=&\inf_{X_i\in\Phi_g} \|X_i\|,\\
R_p=&\inf_{X_i\in\Phi_p} \|X_i\|,
\end{align}
respectively. The cumulative distribution function (CDF) and probability distribution function (PDF) for $ R_p $ are given by 
\begin{align}
&\bP({R_p}\leq r)=1- \exp(-\pi \lambda_p r^2)\label{eq:CDP},\\
&f_{{R_p}}(r)=2\pi\lambda_p r \exp(-\pi\lambda_p r^2).\label{eq:nearestPPP}
\end{align}
since $ \Phi_p $ is a homogeneous Poisson point process. 
\begin{lemma}\label{L:1}
	The CDF of $ R_g $ is given by 
\begin{equation}
	\bP(R_g<r)=\begin{cases}
		0 &\textrm{if } r<0\\
		\frac{\pi r^2}{ s^2} &\textrm{if } 0 \leq r \leq \frac{s}{2},\label{eq:CDG}\\
		2\frac{r^2}{s^2}\left(\frac{\pi-4\arccos\left(\frac{s}{2r}\right)}{2}-2\right)+\sqrt{\frac{4r^2}{s^2}-1} &\textrm{if }
		\frac{s}{2} < r \leq \frac{s}{\sqrt{2}},\\
		1 &\textrm{if } r > \frac{s}{\sqrt{2}},
	\end{cases}
\end{equation}
and its PDF is given by
\begin{align}
f_{R_g}(r)=\begin{cases}
2\pi rs^{-2}  &\textrm{if }0\leq r \leq \frac{s}{2},\label{3}\\
2\pi r s^{-2}- \frac{8r}{s^2}\arccos(\frac{s}{2r}) &\textrm{if }\frac{s}{2}< r \leq \frac{s}{{2}}\sqrt{2},\\
0 &\textrm{otherwise, }
\end{cases}
\end{align}
where the constant $ s $ denotes the width of square lattice, which is equal to $ \frac{1}{\sqrt{\lambda_g}}$.
\end{lemma}
\begin{IEEEproof}
	Appendix \ref{A:0}
\end{IEEEproof}

	The distance from the origin to the nearest point of $ \Phi $ is described by
	\begin{equation}\label{16}
	\bP({R\leq r})=1-(1-\bP(R_p \leq r))(1-\bP(R_g\leq r)),
	\end{equation}
	where $ \bP(R_g<r) $ and $ \bP(R_p<r) $ are given by \eqref{eq:CDP} and \eqref{eq:CDG}, respectively. 


\subsection{Voronoi Tessellation and Average Fraction of Typical Voronoi Cells}

\begin{corollary}\label{corollary:1}
	Let $ \bar{\mathcal{V}_{p}} $ and $ \bar{\mathcal{V}_{g}}  $ denote the average area for the typical cells w.r.t. $ \Phi_p $ and $ \Phi_g $, respectively. The average area fractions are  given by 
	\begin{align}
		\frac{\bar{\mathcal{V}_{p}}}{\bar{\mathcal{V}}_p+\bar{\mathcal{V}}_g}&=\int_{0}^{\sqrt{2}}(1-\frac{\pi}{4} r^2)f(r)\,dr+\int_{1}^{\sqrt{2}}\left(r^2 \arccos(1/r)-\sqrt{r^2-1}\right) f(r) \diff r,\\
	\frac{\bar{\mathcal{V}_{g}}}{\bar{\mathcal{V}}_p+\bar{\mathcal{V}}_g}&=1-\frac{\bar{\mathcal{V}_{p}}}{\bar{\mathcal{V}}_p+\bar{\mathcal{V}}_g},
		\end{align}
	where the function $ f(r) $ is given by 
	\begin{align}
	f(r)= 2 \pi\frac{\rho_\lambda}{4} r\exp\left(-\pi \frac{\rho_\lambda }{4} r^2\right).
	\end{align}
\end{corollary}
\begin{IEEEproof}
	The Voronoi cell centered around $ X^\star $ is described by
	\begin{align*}
	\mathcal{V}_{X^\star}&=\{x\in\bR^2| \|x\|\leq \|x-X_j\|\forall  X_j\in\Phi \}.
	\end{align*}
	The area fraction of the typical cell is then 
\begin{align*}
	 \frac{\bar{\mathcal{V}_{p}}}{\bar{\mathcal{V}}_p+\bar{\mathcal{V}}_g}&=\frac{\bE^{0}(\mathbbm{1}\{R_g>R_p\}}{\bE^{0}(\mathbbm{1}\{R_g>R_p\}+\bE^{0}(\mathbbm{1}\{R_p>R_g\})}\\
	 &=\bE^{0} \left(\mathbbm{1}\left\{R_g>R_p \right\}\right)\\
	 &=\int_{0}^\infty\bP\left(R_g\geq{r}\right)f_{R_p}(r) \diff r.\nnb
	 \end{align*}
	 By replacing $ \bP(R_g>r) $ with expression \eqref{eq:CDG}, we have 
	 \begin{align*}
	 \frac{\bar{\mathcal{V}_{p}}}{\bar{\mathcal{V}}_p+\bar{\mathcal{V}}_g}&=\int_{0}^{\sqrt{2}}(1-\frac{\pi}{4} r^2)f(r)\,dr+\int_{1}^{\sqrt{2}}\left(r^2 \arccos(1/r)-\sqrt{r^2-1}\right)f(r) \diff r,
\end{align*}
where $ 	f(r)= 2 \pi\frac{\rho_\lambda}{4} r\exp\left(-\pi \frac{\rho_\lambda }{4} r^2\right). $
\end{IEEEproof}

\subsection{Application to Any Pairwise Repulsive Network}\label{Preliminaries}
The applicability of the network model is supported by discussing the pair correlation function of  proposed point process. It is known that the pair correlation function is a fundamental mathematical metric {for explaining the pairwise repulsion between two points} \cite[Section 4.5.]{chiu2013stochastic}. The concept was utilized in cellular architecture by the  authors of \cite{taylor2012pairwise}. 
\begin{definition}
	Consider a simple point process $ \Xi $. The pair correlation function between $ x,y\in\Xi $ is defined by
	\begin{equation}
	\kappa(x,y)=\frac{\lambda(x,y)}{\lambda(x)\lambda(y)},
	\end{equation}
	where $ \lambda(x,y) $ is the factorial second-order moment measure of $ \Xi $, and $\lambda(x)$ is the first-order moment measure of $ \Xi $ at $ X $. 
\end{definition}
To describe the pair correlation function of our proposed model $\Phi$, we define two sets $D=\{ (x,y): x\neq y \in\mathbb{R}^2 \}$ and $G=\{ (x,y): x\sim_g y\}$ where $x\sim_g y$ means $x-y\in s\cdot k$ for some $k\in\mathbb{Z}^2$. In other words, $D$ is a collection of all diagonal points in $\mathbb{R}^2\times \mathbb{R}^2$, and $G$ is a collection of all possible pair grid points from each realization of $\Phi_g$. 
\begin{proposition}\label{prop:pair_correlation}
	The pair correlation function of the proposed $ \Phi $ is given by
	\begin{equation}\label{eqn_kappa}
	\kappa(x,y){=} 1-\frac{1}{\left(1+\rho_\lambda\right)^2}  \quad {\forall (x,y)\in \mathbb{R}^2\times\mathbb{R}^2 \setminus \left(D \cup G\right),} 
	\end{equation} 
	which is independent of $ x,y $. Equivalently,
	\begin{equation}\label{rhoramda}
	\rho_\lambda=\frac{1}{\sqrt{1-\kappa(x,y)}}-1 \quad {\forall (x,y)\in\mathbb{R}^2\times\mathbb{R}^2 \setminus \left(D \cup G\right).}
	\end{equation}
\end{proposition}
\begin{IEEEproof}
	To find the factorial second moment density function $\lambda(x,y)$ for $(x,y)\in \mathbb{R}^2\times\mathbb{R}^2 \setminus \left(D \cup G\right)$, we can consider three sub-cases where each point can be from either the grid or the PPP excluding both from the grid. Since $\Phi_p$ and $\Phi_g$ are independent,
	\begin{equation}
	\lambda(x,y)=\lambda_p^2+\lambda_p\lambda_g+\lambda_g\lambda_p. 
	\end{equation}	
	Then by using $\lambda(x)=\lambda(y)=\lambda_p+\lambda_g$ for all $x,y\in\mathbb{R}^2$, \eqref{eqn_kappa} follows immediately. 
\end{IEEEproof}
\par Since $D\cup G$ has zero volume in $\mathbb{R}^2\times \mathbb{R}^2$, we can regard the pair-correlation function $\kappa(x,y)$ of $\Phi$ as a constant function $\kappa(x,y)=\kappa$ for all $x,y\in \mathbb{R}^2$. 
In general, for any simple point process $\Xi$, the pair correlation function explains the pairwise repulsion between two points. Since we have the one-to-one relation between $ \rho_\lambda $ and $ \kappa$, our proposed point process $\Phi$ is able to capture pairwise repulsion by a single scalar constant. In the following example, we illustrate one method that the proposed model and its results can be applied to a spatial model $ \Xi $ with arbitrary BSs utilizing the average pair correlation $\hat{\kappa}$ in \cite[Section 4.7.4]{chiu2013stochastic}.  
\begin{example}
	\textbf{The actual BS deployment map and the applicability of our model}: Figures \ref{Data1} and \ref{Data2} demonstrate two actual BS deployment maps of the United States at latitude and longitude specified by $ [33.45,33.55]\times [-112.2,-112.05] $ and $ [33.34,33.44]\times[-112.2,-112.05] $, respectively. We use a open source software \texttt{R} to find the averaged pair correlation values given by 0.90 and 0.766, respectively. Using \eqref{rhoramda}, the intensity ratios of those maps are given by 2.3 and 1.11, respectively. 
\end{example}

\section{User Association, Interference, and Coverage Probability}\label{s:4} 
Throughout this paper, we {write} $ \cA_g$ and $\cA_p $ to denote the event that the typical user being associated with either $ \Phi_g $ or $ \Phi_p $, respectively.  
\subsection{Typical User Association}
\begin{theorem}\label{L:2}
	The probability that the typical user is associated with the random BS is
	\begin{align}
	\bP^0(\cA_p)&=\int_{0}^{\sqrt{2}}(1-\frac{\pi}{4} r^2)f(r)\,dr+\int_{1}^{\sqrt{2}}\left(r^2 \arccos(1/r)-\sqrt{r^2-1}\right) f(r) \diff r.\label{eq:T1-0}
	\end{align}
	Accordingly, the probability that the typical user is associated with the grid BS is
	\begin{equation}
	\bP^0(\cA_g)=1-\bP^0(\cA_p)\label{eq:T1-01},
	\end{equation}
	where $ f(r) $ is given by 
	\begin{align}
	f(r)= 2 \pi\frac{\rho_\lambda{\eta}^{2/\alpha}}{4} r\exp\left(-\pi \frac{\rho_\lambda \eta^{2/\alpha}}{4} r^2\right).
	\end{align}
	
\end{theorem}
\begin{IEEEproof}
	Appendix \ref{A:0-1}
\end{IEEEproof}
\par Note that the expressions given in \eqref{eq:T1-0} or \eqref{eq:T1-01} are functions of the intensity ratio $ \rho_\lambda $ and power ratio $ \eta$. Note that if  $ \eta=1$ the association probability reduces to the average area fraction provided in Corollary \ref{corollary:1}.  In the following, we derive the upper and lower bounds for the association probability. 

\begin{corollary}\label{P:2}
	The association probability $ \bP^0(\cA_p) $ is lower and upper bounded by 
\begin{align}
\bP^0(\cA_p)&\geq1+\frac{e^{-\pi \rho /2}-1}{\rho}+ \frac{\pi/2-1}{\sqrt{\rho}(\sqrt{2}-1)}\beta-\gamma\label{eq:P-1-1},\\
\bP^0(\cA_p)&\leq1+\frac{e^{-\pi \rho /2}-1}{\rho}+ \frac{\pi/2-1}{\sqrt{\rho}(\sqrt{2}-1)}\beta\label{eq:P-1-2},
\end{align}
where $ \rho=\rho_\lambda{\eta}^{2/\alpha}\label{rhoisrho} $,  $ \erf(z)=\dfrac{2}{\sqrt{\pi}}\int_{0}^z e^{-t^2}\diff t $,
\begin{align}
\beta&=\erf(\sqrt{\pi\rho/2})-\erf(\sqrt{\pi\rho/4}),\\
\gamma&=0.0956(e^{-\pi\rho/4}-e^{-\pi\rho/2}).
\end{align}
\end{corollary}
\begin{IEEEproof} Appendix \ref{AA}.
\end{IEEEproof}
\par
\begin{figure}
\centering
\includegraphics[width=0.8\linewidth]{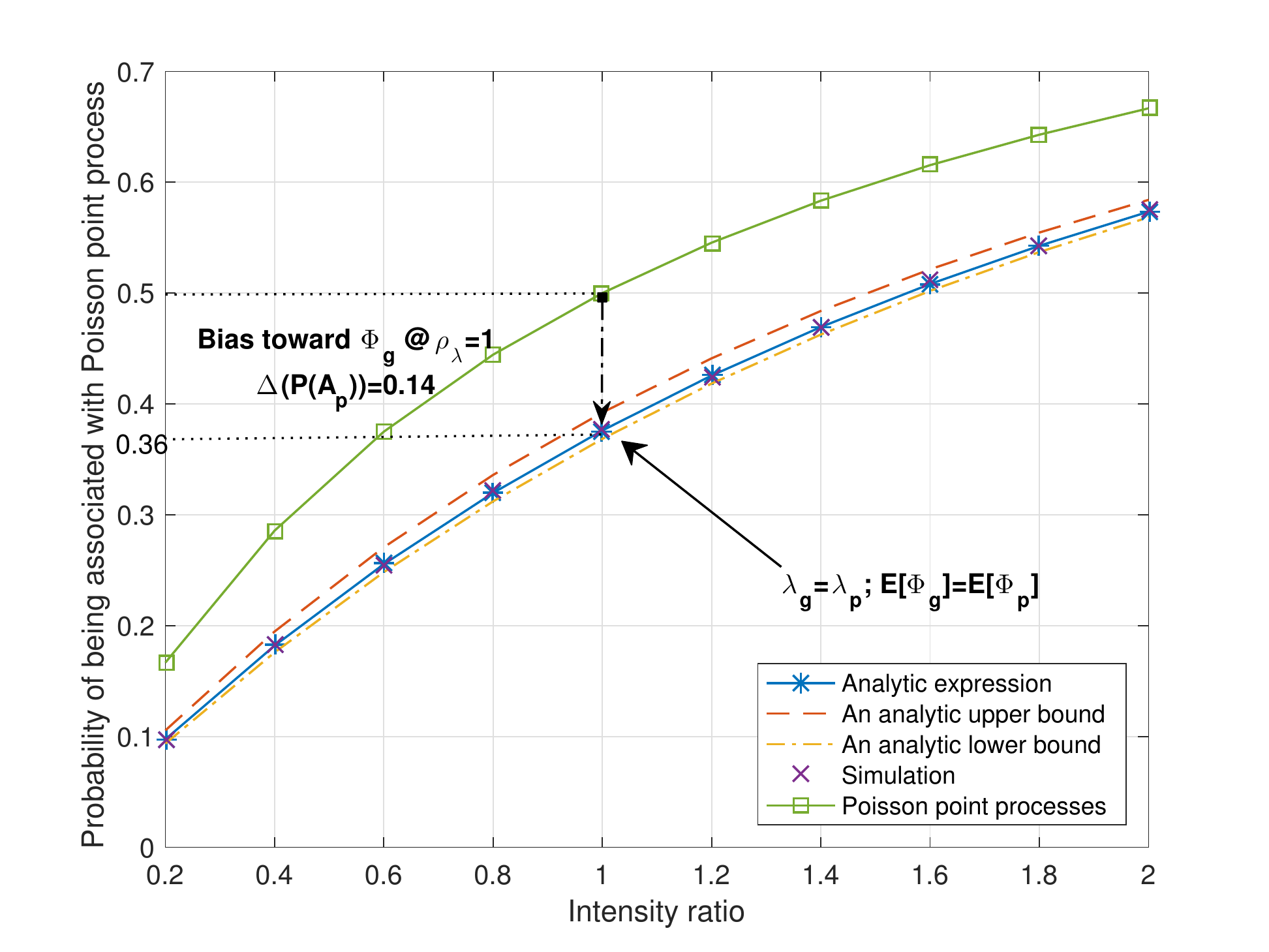}
\caption{$ \bP(\mathcal{A}_p) $ (square mark) is juxtaposed with $ \bP(\mathcal{A}_{p'}) $ (x mark) for $ \rho_\lambda\in(0.2,2) $. For $ \rho_\lambda=1 $, the proposed model with repulsion  gives  $ \bP(\cA_p)=0.36 $  while the model with only PPPs produces $ \bP(\cA_p)=0.5 $.}
\label{fig:1}
\end{figure}

 Figure \ref{fig:1} demonstrates the association probability, its simulated value, and its lower and upper bounds. The lower and upper bounds are very tight. The bounds are described very simple function will show its best use to study the user association in repulsive networks.
\begin{remark}
To illustrate the influence of repulsion to user association, let us consider the same transmit power for grid BSs and random BSs for the moment. In this case, for   $\lambda_p=\lambda_g $, 
 $$ \bP(\cA_p)=0.36 \text{ and } \bP(\cA_g)=0.64. $$  In other words, our model with repulsion yields
$ \bP^0(\cA_p)< \bP^0(\cA_g).
  $ This contrasts to the user association in cellular networks only modeled by independent PPPs. Specifically,  the association probability of the typical user in the independent PPP BSs $ \Phi_p $ and $ \Phi_{p'} $ with intensity $ \lambda_p $ and $ \lambda_{p'} $ was described by 
\begin{align}
\bP^0({\cA_p})=\frac{\lambda_{p}}{\lambda_{p'}+\lambda_p} \text{ and }
\bP^0({\cA_{p'}})=\frac{\lambda_{p'}}{\lambda_{g}+\lambda_p}\label{eq:poisson}.
\end{align} 
In this case, for $ \lambda_p=\lambda_{p'} $, 
\begin{equation}
 \bP^0({\cA_p})=0.5 \text{ and }\bP^0({\cA_{p'}})=0.5. \label{samepp}
\end{equation} 
In other words, by comparing the proposed model to the model with only PPPs, even for the same intensity, the typical user is more likely to be associated with the grid BSs. Figure \ref{fig:1} shows $ \bP^0(\cA_p) $ in the proposed model and in the model with only PPPs.  
The difference of association probabilities is evident and it is alluded by the deterministic structure of the model. Specifically, in the network model with the grid structure, a {random user} finds its associated BS at a distance always less than infinity; $ \frac{1}{2\sqrt{\lambda_g}} $ in this paper. In other words, the void probability of the proposed point process has a finite support, $ (0,\frac{1}{2\sqrt{\lambda_g}}) $. However, the void probability of the models with only PPPs has an \emph{infinite} support, {$ (0,\infty) $}. 
\end{remark} 

\subsection{Coverage Probability under Exponential Fading}
To elaborate the derivation of the coverage probability, consider two PPPs denoted by $ \Phi_p $ and $ \Phi_{p'} $. The SIR coverage probability is given 
	\begin{align}
	\bP^0(\SIR>T)&=\bP^0\left( \SIR > T , \mathcal{A}_{p'} \right)  + \mathbf{P}^0\left( \SIR > T , \mathcal{A}_p \right)\nnb\\
	&=\bP^0(\SIR>T|\cA_p')\bP^0(\cA_p')+\bP^0(\SIR>T|\cA_p)\bP^0(\cA_p)\label{eq:40}, 
	\end{align}
	where sometimes produces a simple closed-form expression \cite{haenggi2009interference}.

On the other hand, in the proposed BS model the complete independence property \cite[Chapter 2.]{daley2007introduction} does not hold and \eqref{eq:40} produces complex expression. In particular, given the association of users, the formula for the interference needs to be derived from conditional point processes such that there are no points of $ \Phi_p $ and $ \Phi_g $ inside a certain exclusion area.

  We use $\Phi_{g}^{!\sim r}$ and $\Phi_{p}^{!\sim r}$ to denote the conditional point processes of $ \Phi_g  $ and $\Phi_p $ such that no points of $ \Phi_g $ and $ \Phi_p $ exist inside the ball of radius $ r $ centered at zero, respectively.
Note that users are assumed to be associated with the BSs with the maximum average receive power. Therefore, provided that the typical user is associated with a grid BS at distance $ \|U\| $, the radius of the exclusion ball for $ \Phi_p $ is given by $ \|U\|\eta^{1/\alpha} $. In the same vein, provided that the typical user is associated with a Poisson BS at the distance $ R $, the radius of the exclusion ball for the grid is given by $ R\eta^{-1/\alpha} $. 
In the following, we omit the Palm notation for simpler expression. The coverage probability is given by
\begin{align}
&\mathbf{P}\left( \SIR > T , \mathcal{A}_g \right) \nnb\\
&= \mathbf{P}\left( \frac{P_{X^{\star}}H_{X^\star}/\|X^\star\|^{\alpha}}{\sum\limits_{X_i\in \Phi\setminus \{X^{\star}\}} P_{X_i}H_{X_i}/\|X_i\|^\alpha} > T ,  R_p>\|X^\star\|\eta^{1/\alpha} \right)\nnb\\
&\stackrel{(a)}{=}\bE_U\left[\bP\left(\left.\frac{P_{X^\star}H_{X^\star}/\|X^\star\|^{\alpha}}{\sum\limits_{X_i\in \Phi\setminus \{X^\star\}} P_{X_i}H_{X_i}/\|X_i\|^\alpha} > T, R_p>\|X^\star\|\eta^{1/\alpha} \right|X^\star\stackrel{d}{=}U\right)\right]\nnb\\
&\stackrel{(b)}{=}\bE_U\left[\bP_{\cdot|U}\left(\left.\frac{P_{U}H_{U}/\|U\|^{\alpha}}{\sum\limits_{X_i\in \Phi\setminus \{U\}} P_{X_i}H_{X_i}/\|X_i\|^\alpha} > T\right| R_p>\|U\|\eta^{1/\alpha} \right)\bP_{\cdot|U}\left( R_p>\|U\|\eta^{1/\alpha}\right)\right]\nnb\\
&\stackrel{(c)}{=}\bE_U\left[\bP_{\cdot|U}\left(\frac{p_gH_{U}/\|U\|^\alpha}{\sum\limits_{X_i\in \Phi_{g}\setminus\{U\}\cup \Phi_{p}^{ !\sim \|U\|\eta^{1/\alpha}}} P_{X_i}H_{X_i}/\|X_i\|^\alpha } > T\right)\bP_{\cdot|U}(R_p>\|U\|\eta^{1/\alpha})\right]\label{eq:xy}
\end{align}
where (a) is given by conditioning on the random variable $ \|U\| $ which is given by \eqref{3}, (b) is obtained by the Bayes rule with the notation $ \bP_{\cdot|U}=\bP(\cdot|U) $, and (c) is acquired from incorporating the exclusion ball. In the expectation of  \eqref{eq:xy}, we have 
\begin{align}
	&\bP_{\cdot|U}\left(\frac{p_gH_{U}/\|U\|^\alpha}{\sum\limits_{X_i\in \Phi_{g}\setminus\{U\}\cup \Phi_{p}^{ !\sim \|U\|\eta^{1/\alpha}}} P_{X_i}H_{X_i}/\|X_i\|^\alpha } > T\right)\nnb\\
	&\stackrel{(d)}{=}\bE_{\cdot|U}\left[\mathbbm{1}\left\{\frac{p_gH_{U}/\|U\|^\alpha}{\sum\limits_{X_i\in \Phi_{g}\setminus\{U\}\cup \Phi_{p}^{ !\sim \|U\|\eta^{1/\alpha}}} P_{X_i}H_{X_i}/\|X_i\|^\alpha } > T\right\}\right]\nnb\\
	&=\bE_{\cdot|U}\left[\bP\left(  H > T\|U\|^\alpha p_g^{-1}\left(I_{\Phi_g\setminus\{U\}}+I_{\Phi_p^{!\sim \|U\|\eta^{1/\alpha}}}\right)\right)\right]\nnb\\
	&\stackrel{(e)}{=}\bE_{\cdot|U}\left[\exp\left(-T\|U\|^\alpha p_g^{-1}\left(I_{\Phi_g\setminus \{U\}}+I_{\Phi_p^{!\sim\|U\|\eta^{1/\alpha}}}\right)\right)\right]\nnb\\
	&=  \mathcal{L}_{I_{\Phi_g \setminus \{U \}} } \left( T \| U \|^{\alpha}p_g^{-1} \right) \mathcal{L}_{I_{\Phi_{p}^{!\sim \|U \| \eta^{1/\alpha}}}} \left( T \|U \|^{\alpha}p_g^{-1} \right) \label{xx},
\end{align} 
	where (d) is given by $ \bP(A)=\bE[\mathbbm{1}\{A\}] $ and (e) is obtained by using CDF of exponential distribution. In \eqref{xx}, $ \cL_{I_{\Phi_g\setminus \{U\}}}(\cdot) $ denotes the Laplace transform of the interference created by $ \Phi_g\setminus \{U\} $. Incorporating \eqref{xx} to \eqref{eq:xy}, we have
	 \begin{align}
	\bP(\SIR>T,\cA_g)&\stackrel{}{=}\mathbf{E}\left[  \mathcal{L}_{I_{\Phi_g \setminus \{U \}} } \left( T \| U \|^{\alpha}p_g^{-1} \right) \mathcal{L}_{I_{\Phi_{p}^{!\sim \|U \| \eta^{1/\alpha}}}} \left( T \|U \|^{\alpha}p_g^{-1} \right)\bP(R_p>\|U\|\eta^{1/\alpha}) \right]\label{eq:49},
\end{align}
where 
\begin{align*}
\bP_{\cdot|U}(R_p>\|U\|\eta^{1/\alpha})&=\bP(R_p>\|U\|\eta^{1/\alpha}|R)=\exp({-\pi\lambda_p\|U\|^2\eta^{2/\alpha}})
\end{align*}
and $ f_{\|U\|}(u)= f_{R_g}(u) $ which is given in \eqref{3}.
\par 
Consequently, the derivation of the coverage expression is reduced into obtaining the conditional expectations of functionals on the conditional point processes: $ {\Phi_g \setminus \{U \}}  $ and $\Phi_{p}^{!\sim \|U \| \eta^{1/\alpha}}. $
\par Similarly, we have 
\begin{align}
&\mathbf{P}\left( \SIR > T , \mathcal{A}_p \right) \nnb\\
&= \mathbf{P}\left( \frac{P_{X^{\star}}H_{X^\star}/\|X^\star\|^{\alpha}}{\sum\limits_{X_i\in \Phi\setminus \{X^{\star}\}} P_{X_i}H_{X_i}/\|X_i\|^\alpha} > T ,  R_g>\|X^\star\|\eta^{-1/\alpha} \right)\\
&=\bE_R\left[\bP\left(\left.\frac{P_{X^\star}H_{X^\star}/\|X^\star\|^{\alpha}}{\sum\limits_{X_i\in \Phi\setminus \{X^\star\}} P_{X_i}H_{X_i}/\|X_i\|^\alpha} > T, R_g>\|X^\star\|\eta^{-1/\alpha} \right|\|X^\star\|\stackrel{d}{=}R\right)\right]\\
&=\bE_R\left[\bP_{\cdot|R}\left(\frac{p_pH_{R}/\|R\|^\alpha}{\sum\limits_{X_i\in \Phi_{p}^{!\sim R}\cup \Phi_{p}^{ !\sim R\eta^{-1/\alpha}}} P_{X_i}H_{X_i}/\|X_i\|^\alpha } > T\right)\bP_{\cdot|R}(R_g>R\eta^{-1/\alpha})\right]\label{eq:xy2}.
\end{align}
Similarly to the above, we have 
\begin{equation}
\bP(\SIR>T,\cA_p)=\mathbf{E}_R\left[  \mathcal{L}_{I_{\Phi_g \setminus R\eta^{-1/\alpha}} } \left( T R^{\alpha}p_p^{-1} \right) \mathcal{L}_{I_{\Phi_{p}^{!\sim R }}} \left( T R^{\alpha}p_p^{-1} \right)\bP_{\cdot|R}(R_g>R\eta^{-1/\alpha})\right]\label{eq:50},
\end{equation}
where 
\begin{align*}
\bP_{\cdot|R}(R_g>R\eta^{-1/\alpha})&=\bP(R_g>R\eta^{-1/\alpha}|R)\\
&=\frac{\nu_2(S_0\setminus B(0,R\eta^{-1/\alpha}))}{\nu_2(S_0)},\\
	F_{R}(r)&=2\pi\lambda_pr \exp({-\pi\lambda_p r^2}).
\end{align*}
Consequently, deriving the coverage probability of the typical user is reduced into obtaining the conditional expectations of the functionals on the conditional point processes, $ {\Phi_g^{!\sim R\eta^{-1/\alpha}} }   $ and ${\Phi_{p}^{!\sim R}} $.
\subsection{Interference Seen at the Typical User}

\par The Laplace transforms of interference of  $ \Phi_{p}^{\sim R}, \Phi_{p}^{\sim \|U\|\eta^{1/\alpha}}, \Phi_g^{!\sim R\eta^{-1/\alpha}} $ and $ {\Phi_g\setminus\! \{U\}} $ are given.
\begin{lemma}\label{Lemma-3}
	The Laplace transforms of interference from $ \Phi_{p}^{\sim R} $ and $ \Phi_{p}^{\sim \|U\|\eta^{1/\alpha}} $ are
	\begin{align}
	\cL_{ I_{\Phi_{p}^{\sim R} }}\left( \xi \right)&= \exp\left(-2\pi\lambda_p\int_{R}^{\infty}\frac{p_p\xi u^{1-\alpha}}{1+p_p\xi u^{-\alpha}}\diff u\right)\label{eq:P5-1},	\\
		\cL_{ I_{\Phi_{p}^{\sim \|U\|\eta^{1/\alpha}} }}\left( \xi \right)&= \exp\left(-2\pi\lambda_p\int_{\|U\|\eta^{1/\alpha}}^{\infty}\frac{p_p\xi u^{1-\alpha}}{1+p_p\xi u^{-\alpha}}\diff u\right) \label{eq:P5-0}.
	\end{align}		
\end{lemma}
\begin{IEEEproof}
Since we assumed a homogeneous PPP, applying the radius of the exclusion ball and the Markov property directly produces the expressions.
\end{IEEEproof}
\begin{lemma}\label{Lemma-4}
	The Laplace transforms of interference from $ \Phi_g^{!\sim R\eta^{-1/\alpha}} $ and $ {\Phi_g\setminus \{U\}} $ are 
	\begin{align}
		\mathcal{L}_{I_{\Phi_g^{!\sim R\eta^{-1/\alpha}}} }(\xi)&=\int_{S_0\setminus B(0,R\eta^{-1/\alpha})}\prod_{(z_1,z_2)\in\mathbb{Z}^2}\frac{1}{1+p_g\xi\|u'+s(z_1,z_2)\|^{-\alpha}}\bP_{U
			'}(\diff u')\label{eq:L4-0},\\
		\mathcal{L}_{I_{\Phi_g\setminus \{U\}}}(\xi)&=\prod_{(z_1,z_2)\in\mathbb{Z}^2\setminus (0,0)}\frac{1}{1+p_g\xi\|U+s(z_1,z_2)\|^{-\alpha}}\label{eq:L4-1},
	\end{align}
	where $ U' \stackrel{d}{=}\textrm{Uniform}[ S_0\setminus B(0,r\eta^{-1/\alpha}) ]$. 
\end{lemma}
\begin{IEEEproof}
Appendix \ref{A:D}.
\end{IEEEproof}
\subsection{Scaling of the Coverage Probability}
\begin{theorem}\label{Theorem2}
	The coverage probability of the typical user is given by

\begin{align}
	p_\textrm{cov}&=\int_{\hat{S_0}}{e^{-\pi\rho_\lambda\|\hat{u}\|^2\eta^{2/\alpha}}}\prod_{\stackrel{(z_1,z_2)\in \bZ^2}{\neq (0,0)}}\frac{1}{1+\frac{T\|\hat{u}\|^\alpha}{\|\hat{u}+(z_1,z_2)\|^\alpha}}e^{\left.-2\pi\rho_\lambda\int_{\|\hat{u}\|\eta^{1/\alpha}}^\infty\frac{T\eta\|\hat{u}\|^\alpha w^{1-\alpha}}{1+T\eta s^{\alpha}\|\hat{u}\|^\alpha w^{-\alpha}}\diff w\right.}\bP_{\hat{U}}(\diff \hat{u})\nnb\\
	&+\int_{0}^{\infty}{\frac{\nu_2(\hat{S_0'})}{\nu_2(\hat{S_0})}}e^{\left.-2\pi\rho_\lambda\int_{\hat{r}}^\infty\frac{T\hat{r}^\alpha w^{1-\alpha}}{1+T\hat{r}^\alpha w^{-\alpha}}\diff w\right.}\!\!\!\!\!\!\!\!\!\int\limits_{\hat{S}_0\setminus B(\hat{r}\eta^{-1/\alpha})}\prod_{(z_1,z_2)}^ {\mathbb{Z}^2}\frac{\bP_{\hat{U}'}(\diff \hat{u}')}{1+\frac{\eta^{-1}T\hat{r}^\alpha}{\|\hat{u}+(z_1,z_2)\|^{\alpha}}}f_{\hat{R_p}}(\hat{r})\diff \hat{r}\label{eq:prop2},
	\end{align}
		where $ \hat{U}'= \textrm{Uniform}[\hat{S_0}'] $ , $ \hat{S_0'}= \hat{S_0}\setminus B(0,\hat{r}\eta^{-1/\alpha}) $, $ \hat{U}=\textrm{Uniform}[\hat{S_0}] $ , $ \hat{S_0}:=[-1/2,1/2]\times[-1/2,1/2] $, and $ f_{\hat{R_p}}(\hat{r})=2\pi\rho_\lambda \hat{r}e^{-\pi \rho_\lambda \hat{r}^2} $. 
	
	
\end{theorem}
\begin{IEEEproof}
Appendix \ref{A:1}
\end{IEEEproof}
\par 

The coverage expression for the typical user is exactly found in Theorem \ref{Theorem2} and it accurately quantifies the influence of repulsion provided that $ \rho_\lambda $ of a network can be accurately estimated. From  \eqref{rhoramda}, the one-to-one relationship between a pair-wise correlation function and the $ \rho_\lambda $ is given. Therefore, in any network topology where the correlations function can be estimated, equation \eqref{eq:prop2} can be utilized to predict the SIR of the typical user.

\begin{remark}
	From Theorem \ref{Theorem2}, we confirm that the proposed model is scalable; the coverage probability of the typical user is invariant w.r.t. the intensity ratio $ \rho_\lambda $. This is similar to the  homogeneous Poisson network \cite{andrews2011tractable} where the coverage probability is invariant w.r.t. intensity $ \lambda $. Note that, however, the path loss function we considered is a distance-based path loss function with a single exponent and the specific path loss function is known to produce the favorable SIR scale invariant result of Poisson cellular networks given in  \cite{andrews2011tractable} because the order of interference and the order of signal are the same as the density increases \cite{baccelli2015scaling}. In practical studies, it is well known that the received signal power is bounded from above. As in \cite{zhang2015downlink}, under a practical dual slope truncated path loss function, the scale invariant property of SIR coverage probability does not hold. In the same vein, the scaling-invariance property in this paper disappears under different path loss functions and we will numerically examine those functions in Section IV-F. 
	
\end{remark}

\subsection{Bounds on Interference and Coverage Expressions}\label{s:b}
In order to maintain the applicability of the proposed model and to address computational issues, bounds for interference and SIR coverage will be considered.
\begin{proposition}\label{lemma:bounds2}
	Consider an arbitrary square truncation window $ W$. The Laplace transform of interference from $ \Phi_g $ conditioning on $ \mathcal{A}_g $ yields
	\begin{align}
	\mathcal{L}_{I_{\Phi_g \setminus \{U \}} } \left( \xi \right) &\geq e^{\left(-\sum_{z_1,z_2\in\bZ^2\setminus(0,0)}\frac{\xi p_g }{{\|U+s(z_1,z_2)\|}^{\alpha}}\right)}\label{eq:L8-0},\\
	\mathcal{L}_{I_{\Phi_g \setminus \{U \}} } \left( \xi \right) &\leq \left(\prod_{(z_1,z_2)\in\bZ^2\cap W}\frac{1}{1+{\xi p_g}{ \|U+s(z_1,z_2)\|^{-\alpha}}}\right)\label{eq:L8-1}.
	\end{align}On the other hand, the Laplace transform of interference from $ \Phi_g $ conditioning on $ \mathcal{A}_p $ yields
	\begin{align}
	\mathcal{L}_{I_{\Phi_g^{!\sim R\eta^{-1/\alpha}}} } (\xi)  &\geq e^{\left.-\frac{2\pi\int_{R\eta^{-1/\alpha}}^{\infty}v\log\left({1+\xi p_g v^{-\alpha}}\right)\diff v}{\nu_2(S_0')}\right.},\label{eq:L8-5}\\
	 \mathcal{L}_{I_{\Phi_g^{!\sim R\eta^{-1/\alpha}}} } (\xi)  &\leq{}\int_{S_0'}\prod_{(z_1,z_2)\in\bZ^2\cap W}\frac{1}{1+\xi p_g \|v+s(z_1,z_2)\|^{-\alpha}}\bP_{U'}(\diff v)\label{eq:L8-6}.
	\end{align}
\end{proposition}
\begin{IEEEproof}
Given in Appendix \ref{A:2}
\end{IEEEproof}
	As the size of window $ \|W\|$ increases, upper bounds \eqref{eq:L8-1} and \eqref{eq:L8-6} become tighter at the expense of a marginal computational complexity. We will see the tightness of upper bounds in Figure \ref{fig:UB}. In order to derive the simplest coverage expression,  $ W=S_0 $ for the moment.
\begin{proposition}\label{Theorem:3}
	The coverage probability of the typical user is lower bounded by
	\begin{align}
	p_\textrm{cov}&\geq \int_{S_0}{e^{-\pi\lambda_p \|u\|^2\eta^{2/\alpha}}}e^{-\sum_{z_1,z_2}^{\bZ^2\setminus(0,0)}\frac{T\|u\|^\alpha  }{{\|u+s(z_1,z_2)\|}^{\alpha}}}e^{-2\pi\lambda_p\int_{\|u\|\eta^{1/\alpha}}^{\infty}\frac{\eta T\|u\|^\alpha v^{1-\alpha}}{1+\eta T\|u\|^\alpha v^{-\alpha}}\diff v}\bP_U{(\diff u)}\nnb\\
	&+\int_{0}^\infty {\frac{\nu_2(S_0')}{\nu_2(S_0)}}e^{\left.-\frac{2\pi\int_{r\eta^{-1/\alpha}}^{\infty}u\log\left({1+\eta^{-1}Tr^{\alpha}  u^{-\alpha}}\right)\diff u}{\nu_2(S_0')}-2\pi\lambda_p\int_{r}^{\infty}\frac{ T r^\alpha u^{1-\alpha}}{1+T r^\alpha u^{-\alpha}}\diff u\right.} f_{R}(r)\diff r\label{eq:T3-1}.
	\end{align}
	The coverage probability of the typical user is upper bounded by
	\begin{align}
	p_\textrm{cov}&\leq \int_{S_0}{e^{-\pi\lambda_p \|u\|^2\eta^{2/\alpha}}}\frac{e^{-2\pi\lambda_p\int_{\|u\|\eta^{1/\alpha}}^{\infty}\frac{\eta T \|u\|^\alpha  v^{1-\alpha}}{1+\eta T \|u\|^\alpha v^{-\alpha}}\diff v}}{1+T }\bP_U{(\diff u)}\nnb\\
	&+\int_{0}^\infty {\frac{\nu_2(S_0')}{\nu_2(S_0)}}e^{-2\pi\lambda_p\int_{r}^{\infty}\frac{Tr^\alpha v^{1-\alpha}}{1+Tr^\alpha v^{-\alpha}}\diff v} \left(\int_{S'}\frac{1}{{(1+\eta^{-1}T r^\alpha \|u\|^{-\alpha} )}}\bP_{U'}(\diff u)\right)f_{R_p}(r)\diff r \label{eq:T3-2}.
	\end{align}
\end{proposition}
\begin{IEEEproof}
	The expressions \eqref{eq:L8-0} and \eqref{eq:L8-1} evaluated at $ \xi=T\|U\|p_g^{-1} $ yield the lower and upper bounds for $ \bP(\SIR>T,\cA_g) $. Similarly, the expressions \eqref{eq:L8-5} and \eqref{eq:L8-6} evaluated at $ \xi=TR_p p_p^{-1} $  produces lower and upper bounds for $ \bP(\SIR>T,\cA_p) $. 
\end{IEEEproof}

		The expression \eqref{eq:T3-2} requires much less computation compared to \eqref{eq:prop2} because we essentially replace the summation of infinite grid interference by the strongest grid interference. Numerical validation in Section \ref{s:n} will justify it. It supports that interference from longer distances becomes much smaller and it allows the approximation of total interference to the strongest interference for the grid BSs. 

\subsection{Numerical Analysis and Validation}\label{s:n}
\begin{figure}
	\centering
	\includegraphics[width=.8\linewidth]{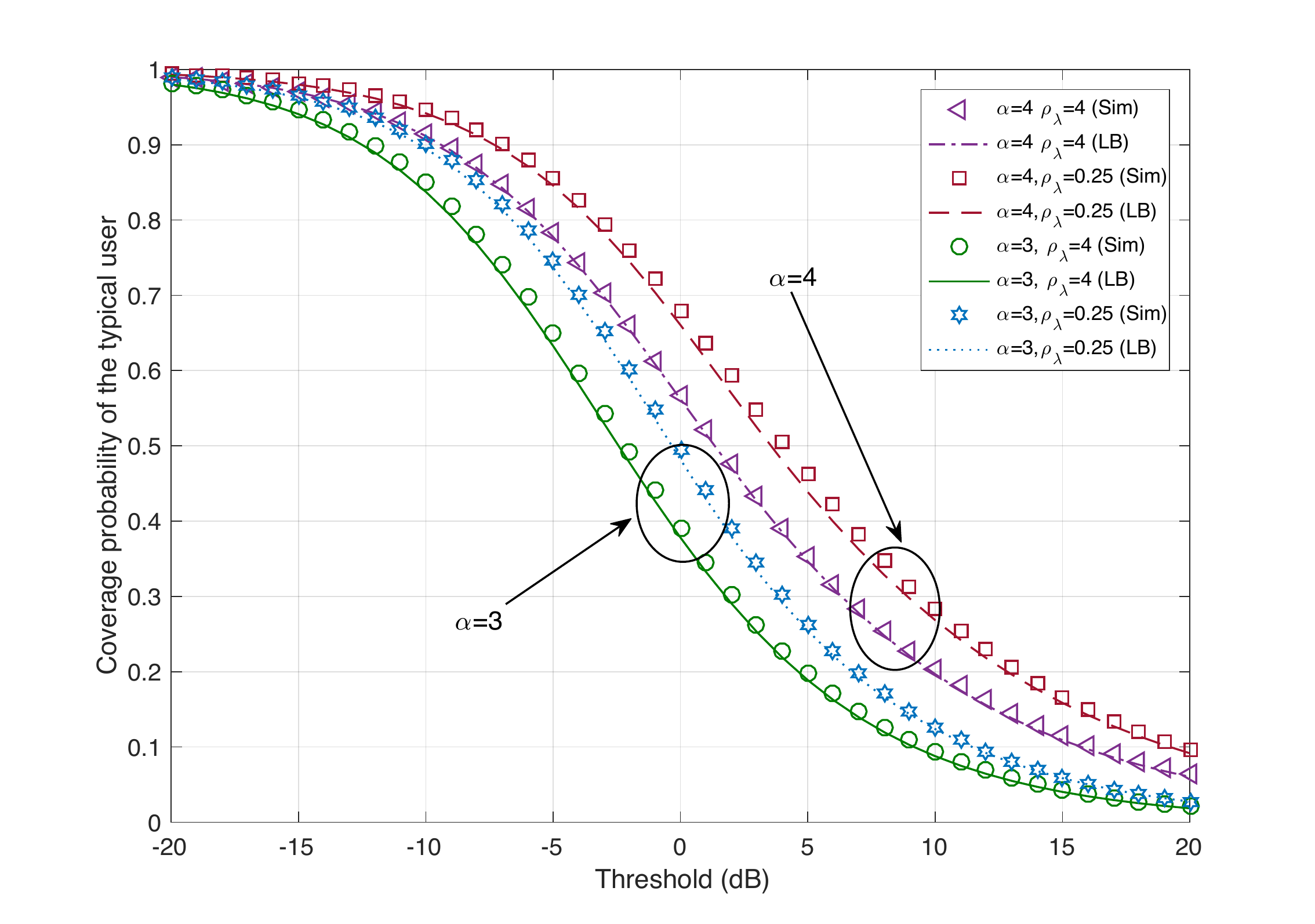}
	\caption{This figure describes the coverage probabilities of the simulation (Sim) and corresponding lower bounds. It delineates $ \alpha=3 ,4 $ and $ \rho_\lambda=0.25,4$. We consider $W=[-3s/2,3s/2]^2 $. To minimize the undesired edge effect, The average number of BSs is maintained to be 100 in the simulation window,  $ \bE[\Phi(K)]=\lambda_g+\lambda_p=100 $ for the multiple values of $ \rho_\lambda$.}
	\label{fig:LB}
\end{figure}

\begin{figure}
	\centering
	\includegraphics[width=0.8\linewidth]{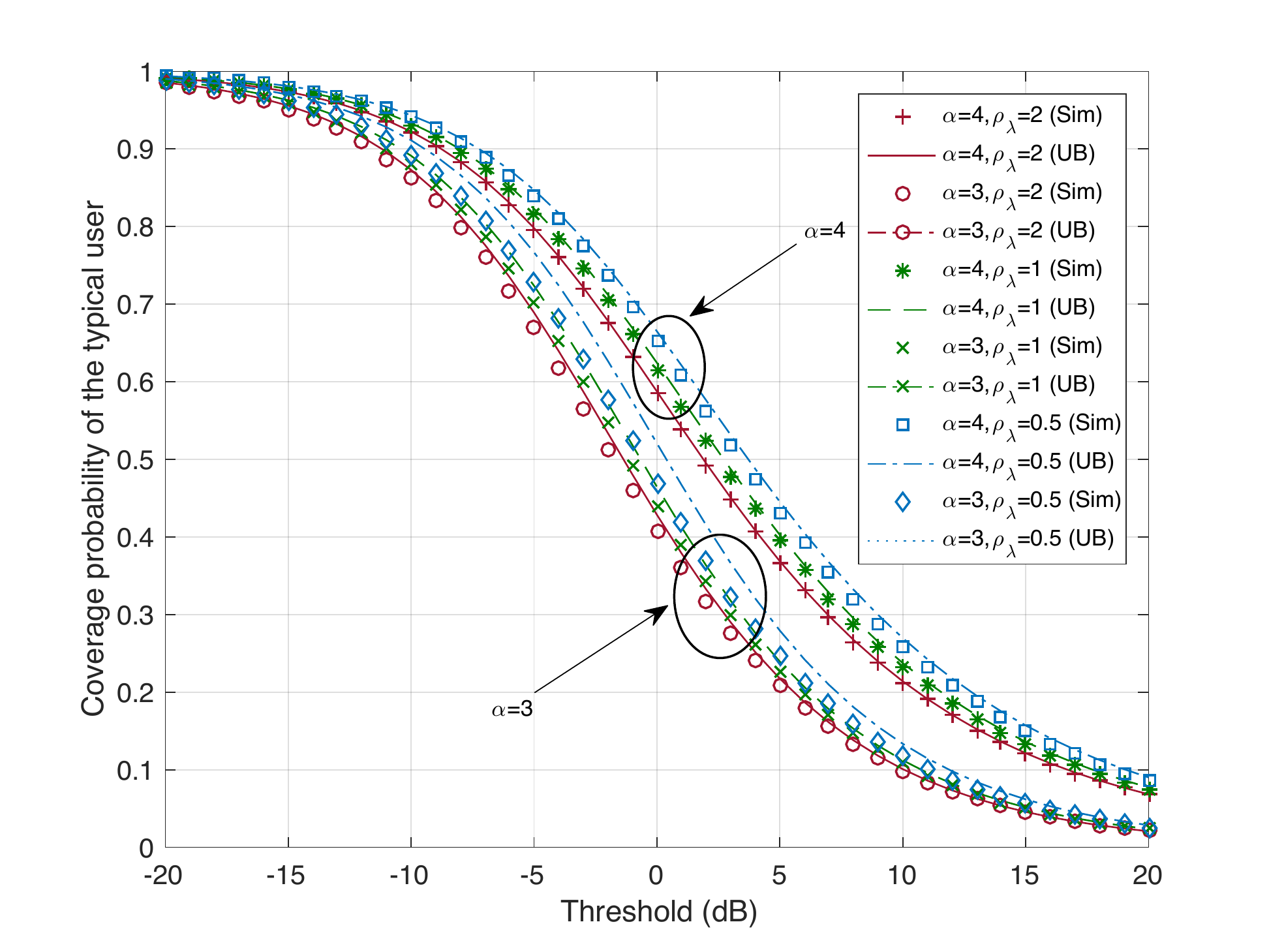}
	\caption{It describes the coverage probabilities of simulations (Sim) and the corresponding upper bounds (UB). It delineates $ \alpha=3, 4 $ and  $ \rho_\lambda=0.5, 2$. We assume $ W=[-3s/2,3s/2]^2. $ To minimize the undesired edge effect, we maintain the average number of BSs in the simulation space at 100,  $ \bE[\Phi(K)]=\lambda_g+\lambda_p=100 $ for the multiple values of $ \rho_\lambda$. }
	\label{fig:UB}
\end{figure}

We compare the coverage probability bounds to the numerically obtained coverage probability of the typical user. We perform the Monte Carlo simulation where the number of iterations is more than $ 10^5 $.   
We depict the coverage probability \eqref{eq:prop2}, the lower bound \eqref{eq:T3-1}, and the upper bound \eqref{eq:T3-2} in Fig. \ref{fig:LB} and \ref{fig:UB}.
	These figures demonstrate that the lower bound and the upper bounds are very tight for a wide range of SIR thresholds. The bounds become tighter as the window size $ W $ increases. The acquired bounds show that interference from further BSs is very small and we can approximate the total interference by the strongest interference.  The variable $ \rho_\lambda $ play a more noticeable role for higher $ \alpha $.  
	
	\par
\begin{figure}
		\centering
		\includegraphics[width=0.8\linewidth]{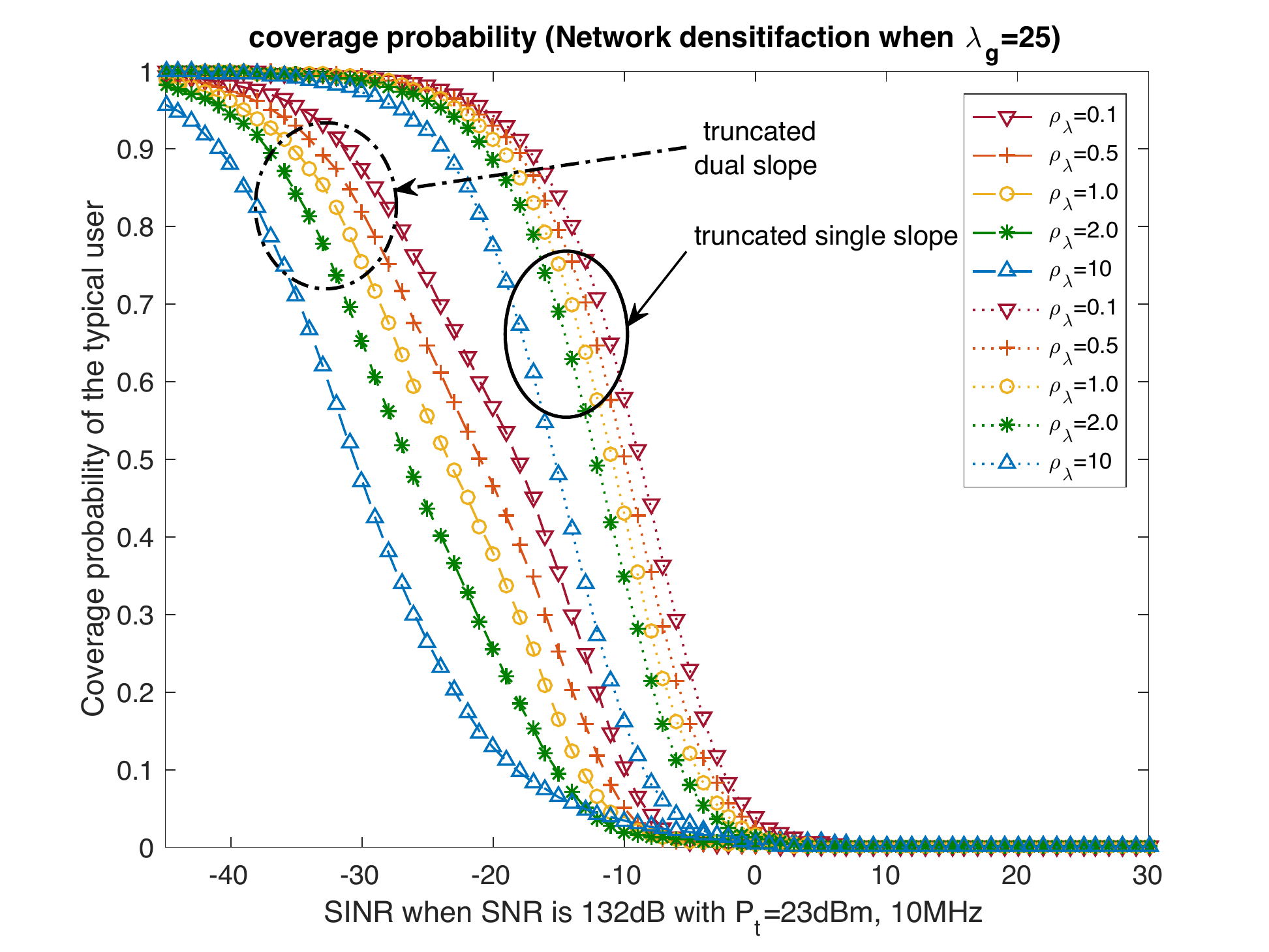}
		\caption{The coverage probability of the typical user where the path loss functions \eqref{los1} and \eqref{los2} are considered. }
		\label{fig:practical}
	\end{figure}

 In Fig. \ref{fig:practical}, the coverage probability of the typical user is investigated under different practical path loss functions  such as
	\begin{align}
	l_1(r)&=\min(C_0,r^\alpha),\label{los1}\\
	 l_2(r)&=\begin{cases}
	 \min(C_0,r^{-\alpha_1}) &\text{ for } 0<r<r_1\\
	 \min(C_1 r^{-\alpha_1}, C_2r^{-\alpha_2}) &\text{ for } r_1<r
	 \end{cases}\label{los2}
	 \end{align} 
	where $ C_0,C_1,C_2,r_1 >0$ are arbitrary constants that ensure the functions continuous.  A more rigorous analysis of coverage probability under different path loss functions is left as  future work. 
\par
\begin{figure}
	\centering
	\includegraphics[width=0.8\linewidth]{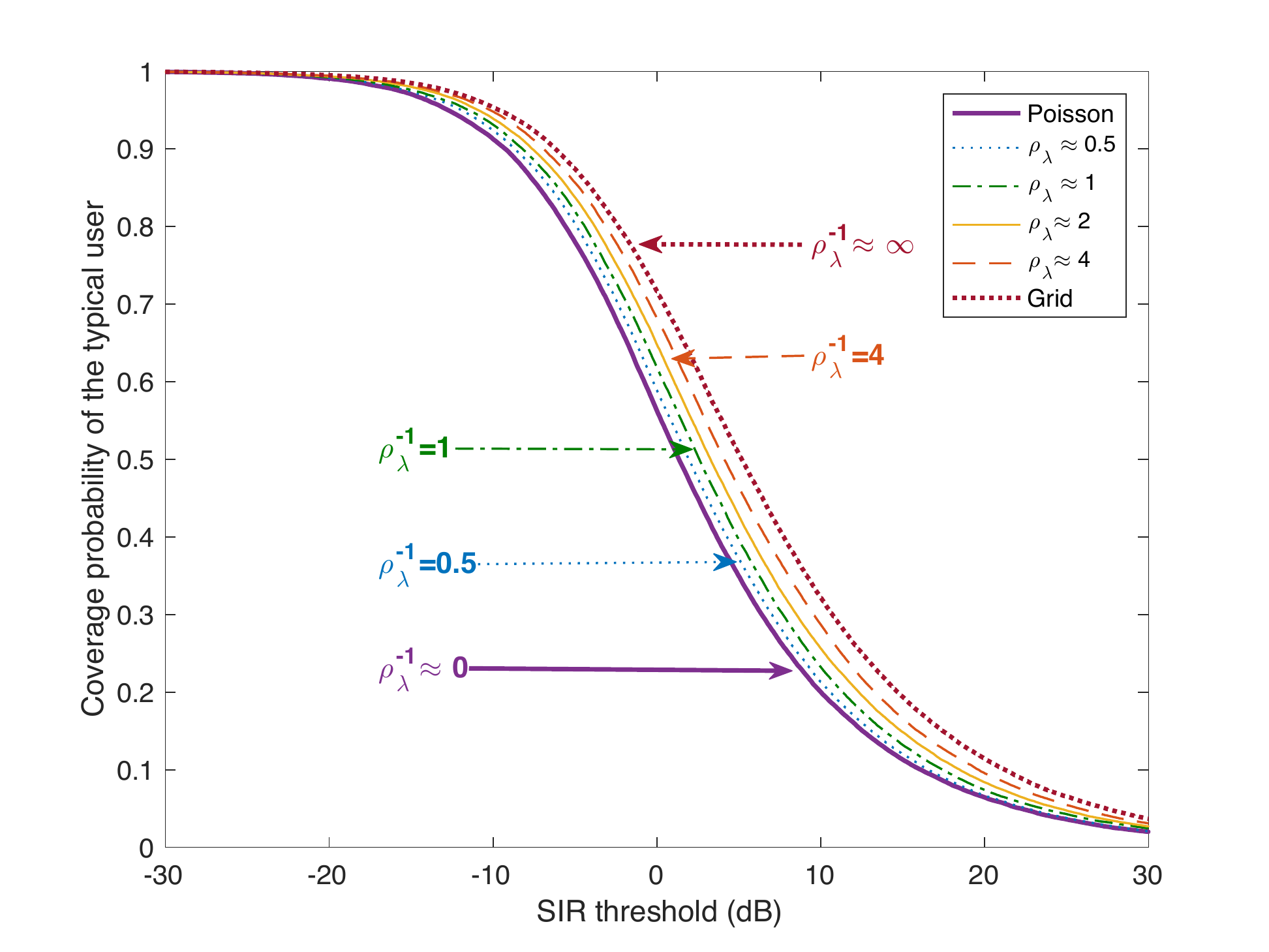}
	\caption{The coverage probability of the typical user increases as the intensity ratio $ \rho_\lambda $ decreases. }\label{Fig:7}	
\end{figure}

\begin{figure}
	\centering
	\includegraphics[width=0.8\linewidth]{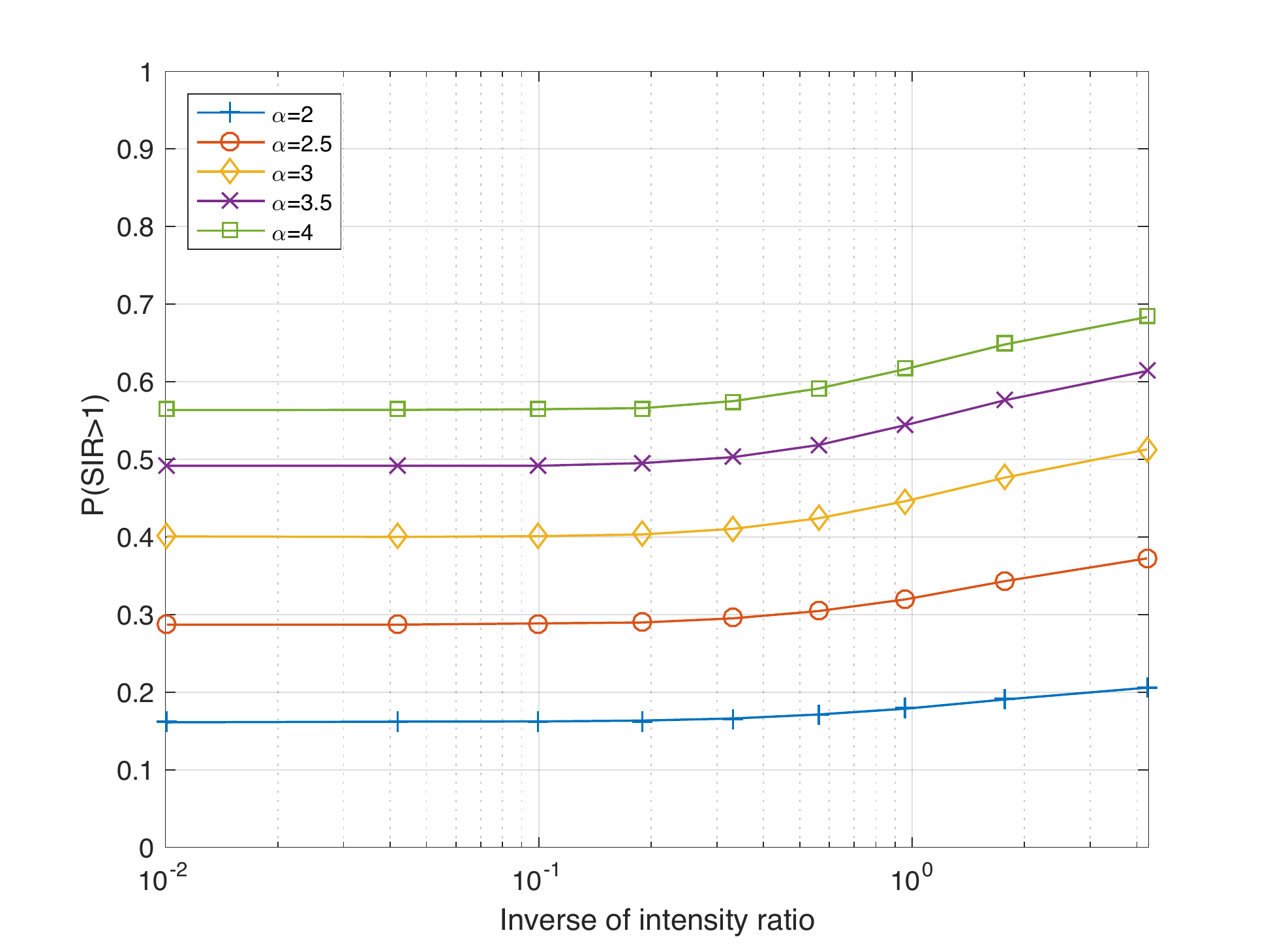}
	\caption{{The typical SIR coverage probability with threshold 1 for multiple values of $ \alpha $.}}\label{Fig:8}	
\end{figure}

 Fig. \ref{Fig:7} illustrates the coverage probability of the typical user for different values of $  \rho_\lambda $. It demonstrates only Poisson $ (\rho_\lambda\approx\infty) $ and only grid BS $ (\rho_\lambda\approx 0) $ for $ \alpha=4 $. The coverage probability of the proposed network is located between that of grid BSs and that of PPP BSs.
\begin{remark}
	In Fig. \ref{Fig:7} and \ref{Fig:8},  the coverage probability is shown to be increasing w.r.t $ \rho_\lambda^{-1} $. The behavior of the typical user in the proposed model contrasts to the behavior of the typical user in Poisson networks \cite{andrews2011tractable} where the SIR does not scale w.r.t. the number of BSs. In other words, if $ \rho_\lambda $ decreases, i.e., more repulsive BSs are added, the SIR increases. This observation corresponds to the remark made in \cite{andrews2011tractable} that the fixed grid is the performance upper bound of the coverage probability. A more rigorous analysis of the growth rate of coverage probability is left for future work.
\end{remark}

%

\subsection{Fitness of the Model: Actual BS Deployment Data and Coverage Probability}

In this section, we conduct an empirical study based on real BS deployment data shown in Fig. \ref{Data1} and \ref{Data2}. For each data set, we estimate the averaged pair correlation value $\kappa$ and the intensity ratio $\rho_\lambda$ as follows.
	\begin{itemize}
		\item{Step 1. The pair correlation function is estimated using open source software \texttt{R}.}
		\item{Step 2. $\rho_\lambda$ is obtained using (22).}
		\item Step 3. The coverage probability of the typical user is obtained using Theorem 2.
\end{itemize}
\par
\begin{figure}
		\centering
		\includegraphics[width=0.8\linewidth]{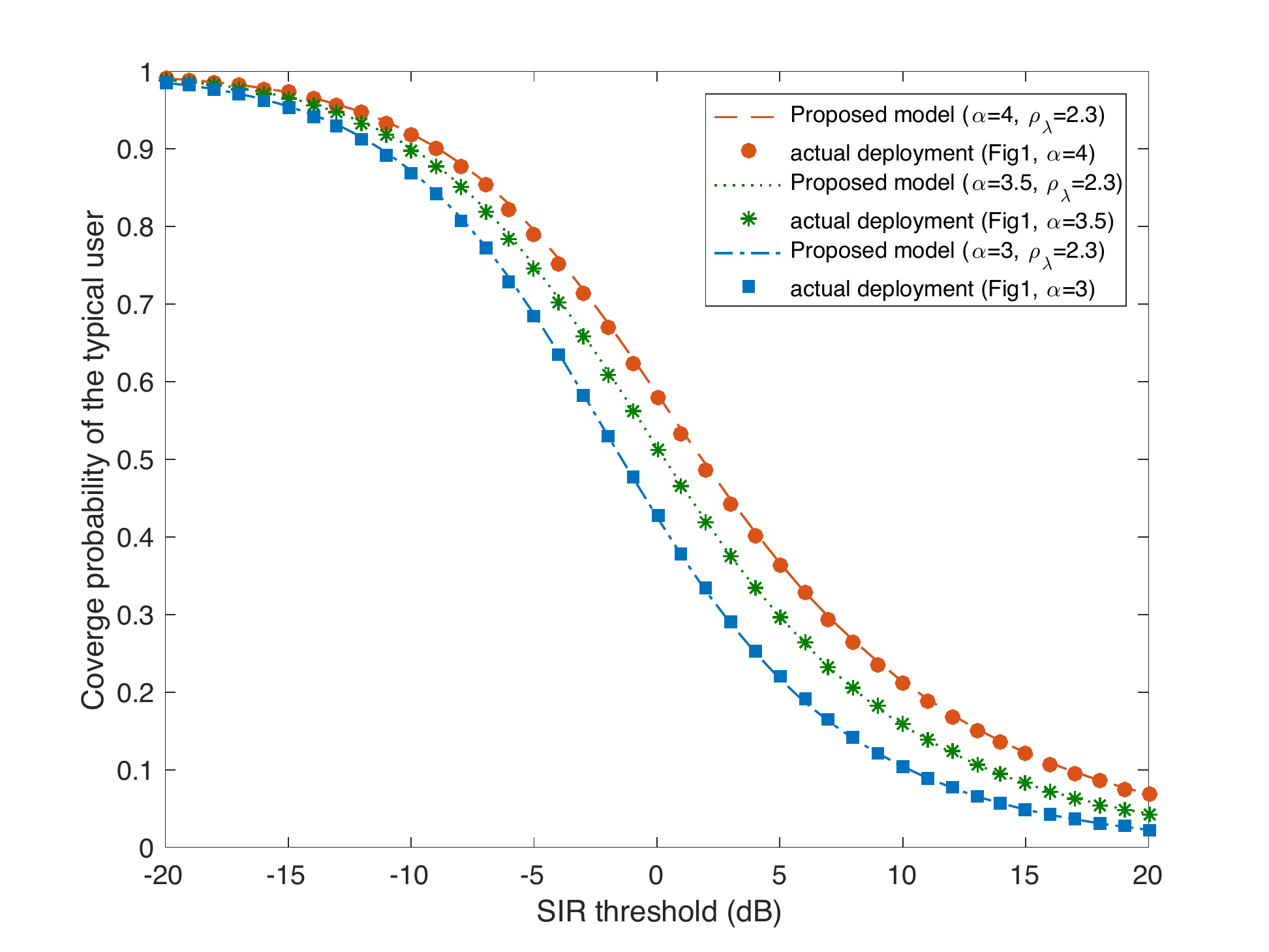}
		\caption{The coverage probability of the actual deployment given in Figure 1 and proposed model with $ \rho_\lambda=2.3 $. The proposed model predicts the actual coverage probability very accurately. Note that the $ \rho_\lambda $ is obtained from physical deployment and the model is robust and consistent for $ \alpha =3,\ 3.5,\ $ and $4.$ }
		\label{fig:coveragereal1}
	\end{figure}

\begin{figure}
		\centering
		\includegraphics[width=0.8\linewidth]{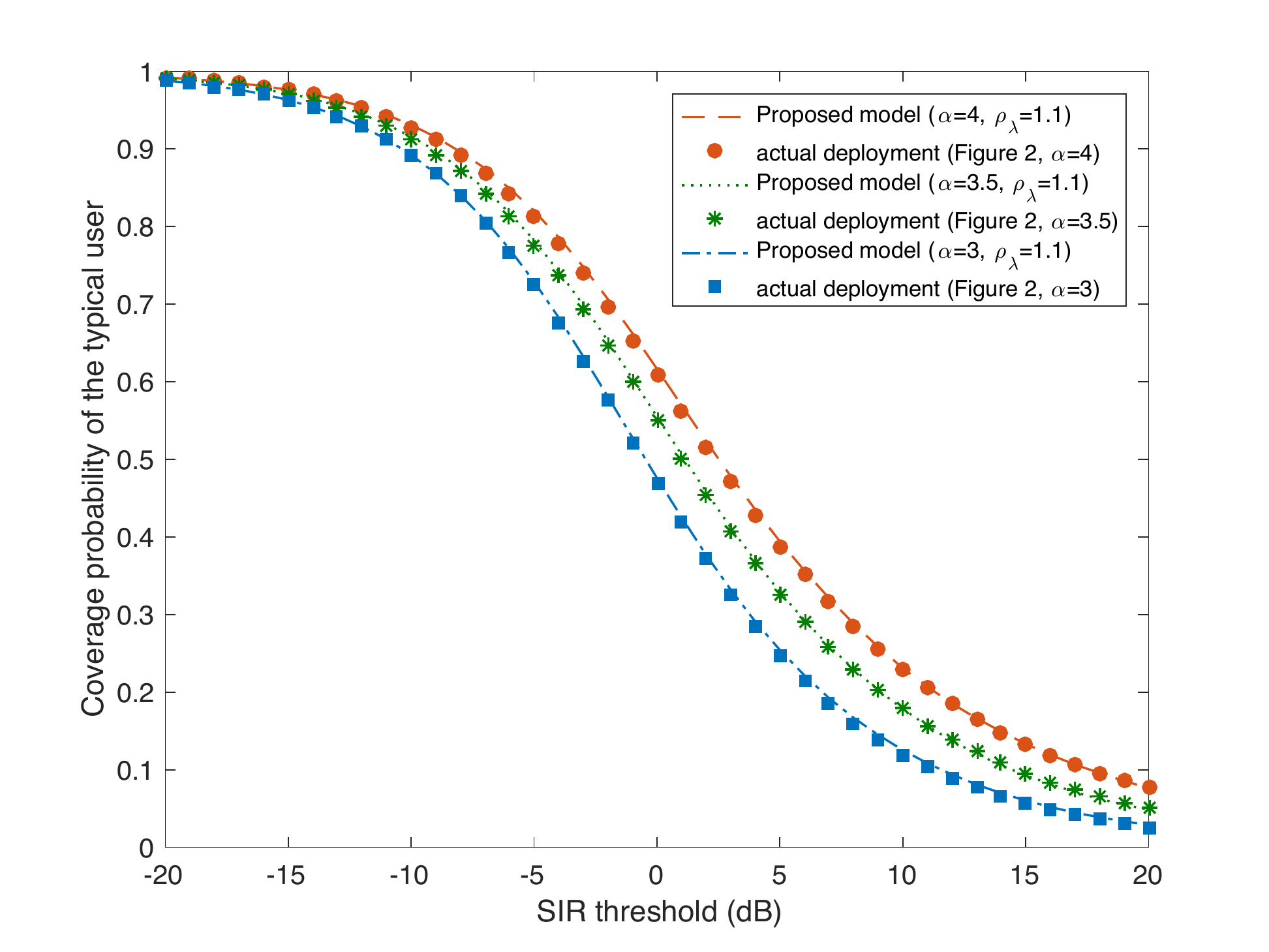}
		\caption{The coverage probability of the actual deployment given in Figure 1 and proposed model with $ \rho_\lambda=1.1 $. The proposed model predicts the actual coverage probability very accurately. Note that the $ \rho_\lambda $ is obtained from physical deployment and the model is robust and consistent for $ \alpha =3,\ 3.5,\ $ and $4.$.}
		\label{fig:coveragereal2}
	\end{figure}

 Fig. \ref{fig:coveragereal1} and \ref{fig:coveragereal2} show the coverage probability obtained from actual deployment and the coverage probability obtained from the proposed model. The figures demonstrate that the proposed model predicts the coverage probability of the actual deployment very accurately. In addition, the parameter $ \rho_\lambda $ of the proposed model is robust and consistent with the change of $ \alpha $. It contrasts to the $ \beta $-Ginibre point process model in \cite{6841639} where the key parameter $ \beta $ should be modified for different path loss exponents; this modification compromises the robustness of the Ginibre model.

\section{Conclusion}\label{s:7}

We developed a scalable modeling technique representing the repulsive BSs and random BSs. We derived the nearest BS distance, the cellular user association probability, the interference seen at the typical user, and the SIR coverage probability. Numerical observations confirmed that the provided bounds for numerical purposes are tight. Our findings include the following three points. First, the typical user is more likely to be associated with a repulsive BS rather than a random BS. Second, we have the expressions for the user association and the coverage  given by $ \rho_\lambda $ and $ \eta $. Finally, the proposed model accurately predicts the coverage probability of the actual cellular deployment.

\appendices
\section{Proof of Lemma \ref{L:1}}\label{A:0}
	Denote a ball of radius $ r $ centered at $ x $ in $ \bR^2 $ by $ B(x,r) $. The nearest distance $ R_g $ is 
\begin{align}
R_g&=\inf\{R>0|\Phi_g(B(0,R))\neq \emptyset\}\nnb\\
&\stackrel{(a)}{=}\inf\{R>0 | \cup_{X_i\in \Phi_g}B(X_i,R)\cap 0 \neq \emptyset \}\label{eq:24}.
\end{align}
where (a) is obtained by using the stationarity of $ \Phi_g $. Let us define a square of width $ s $ centered at the origin \begin{equation}
S_0= \left(-\frac{s}{2},\frac{s}{2}\right)\times \left(-\frac{s}{2},\frac{s}{2}\right).
\end{equation}
Then, the CDF of $ R_g $  is exactly obtained by
\begin{align}
\bP(R_g<r)&=\bP(\inf\{R>0 | \cup_{X_i\in \Phi_g}B(X_i,R)\cap 0 \neq \emptyset \}<r)\nnb\\
&\stackrel{(b)}{=}\frac{\nu_2 (  S_0\cap B(0,r)  ) }{\nu_2 (S_0)}\label{phigstationary}\\
&\stackrel{(c)}{=}\begin{cases}
0 &\textrm{if } r<0\\
\frac{\pi r^2}{ s^2} &\textrm{if } 0 \leq r \leq \frac{s}{2}\\
2\frac{r^2}{s^2}\left(\frac{\pi-4\arccos\left(\frac{s}{2r}\right)}{2}-2\right)+\sqrt{\frac{4r^2}{s^2}-1} &\textrm{if }
\frac{s}{2} < r \leq \frac{s}{\sqrt{2}},\\
1 &\textrm{if } r > \frac{s}{\sqrt{2}},
\end{cases}
\end{align}
where $ \nu_2 (A)$ denotes the area of set $ A$. We have (b) from the fact that the CDF of the first touch distance to a stationary point process is equivalent to the volume fraction of the stationary point process \cite[Section 6.3.1]{chiu2013stochastic} and have (c) by merely computing the area. Taking the derivative of \eqref{eq:CDG} w.r.t. $ r $ completes the proof. 

\section{Proof of Theorem \ref{L:2}}\label{A:0-1}
 	If the typical user is associated with the nearest BS of the PPP at distance $ R_p $, it implies that received power from the nearest PPP is greater than the received power from the nearest grid. The event $ \cA_p $ is given by
\begin{equation}
\cA_p=\left \{\frac{p_p}{R_p^\alpha}> \frac{p_g}{R_g^\alpha}\right \},
\end{equation}
whereas the event $ \cA_g $ is given by
\begin{equation}
\cA_g=\left \{\frac{p_p}{R_p^\alpha}< \frac{p_g}{R_g^\alpha}\right \}\label{eq:A_g event}=\cA_p^{c}
\end{equation}
since  $U= \cA_p\cup \cA_g$. Denoting the indicator function by $ \mathbbm{1}\{\cdot\} $, we have 
\begin{align}
\bP^0(\cA_p)&=\bE^0\left[\mathbbm{1}\left \{\frac{p_p}{R_p^\alpha}> \frac{p_g}{R_g^\alpha}\right \}\right]\label{aa}\\
&=\int_{0}^\infty\bP\left(R_g\geq{r\eta^{-1/\alpha}}\right)f_{R_p}(r) \diff r\nnb\\
&\stackrel{(a)}{=}\int_{0}^\infty \frac{\nu_2(S_0\setminus B(0,{r\eta^{-1/\alpha}}))}{\nu_2(S_0)}2\pi\lambda_p r e^{-\pi\lambda_p r^2} \diff r\nnb\\
&\stackrel{}{=}\int_{0}^{\frac{s}{2}\eta^{1/\alpha}}(s^2-\pi r^2\eta^{-2/\alpha})2\pi\lambda_p r e^{-\pi\lambda_p r^2}\diff r\nnb\\
&+\int_{\frac{s}{2}\eta^{1/\alpha}}^{\frac{s}{\sqrt{2}}\eta^{1/\alpha}}\left(s^2-4rs\sin(\theta)-2r^2\left(\frac{\pi}{2}-2\theta\right)\right)2\pi\lambda_p r e^{-\pi\lambda_p r^2}\diff r\nnb\\
&\stackrel{(b)}{=}\int_{0}^{\sqrt{2}}(1-\frac{\pi}{4} r^2)f(r)\,dr+\int_{1}^{\sqrt{2}}\left(r^2 \arccos(1/r)-\sqrt{r^2-1}\right)f(r) \diff r,\label{13.1}
\end{align}
where  (a) is given from \eqref{phigstationary} and (b) is given by  change of variables. From equation \eqref{13.1}, $ \bP^0(\mathcal{A}_g) $ is obtained.

\section{Proof of Corollary \ref{P:2}}\label{AA}
We develop the bounds by obtaining lower and upper bounds for the second integration in \eqref{eq:T1-0}. We {consider}
\begin{equation}
g(r)=r^2 \arccos(1/r)-\sqrt{r^2-1}\nnb,
\end{equation}
for $1 \leq r \leq \sqrt{2}$. One can easily check that $ g(r) $ is a differentiable function w.r.t $ r $ and that $ g(r) $ is a convex function since its second derivative w.r.t $ r $ is greater than zero for the interval $ (1,\sqrt{2}) $. Then, by fully exploiting the differentiability and the convexity of $ g(r) $, we propose two linear bounds
\begin{align}
&g(r) \geq \frac{g(\sqrt{2})-g(1)}{\sqrt{2}-1}(r-1.19) +0.1662,\label{eq:42}\\
&g(r) \leq \frac{g(\sqrt{2})-g(1)}{\sqrt{2}-1}(r-1)\label{eq:42-1},
\end{align}
for $ 1\leq r \leq \sqrt{2} $. We obtain the lower bound by applying the mean value theorem to the interval $ (1,\sqrt{2}) $ by approximating the tangential line numerically. We obtain the upper bound by finding the smallest linear function greater than $ g(r) $. 
Replacing $ g(r) $ of \eqref{13.1} with \eqref{eq:42} and \eqref{eq:42-1}, we acquire the lower and upper bounds, respectively.  
\section{Proof of Lemma \ref{Lemma-4}}\label{A:D}
For the Laplace transform of the interference from the grid, we have 
\begin{align}
\mathcal{L}_{I_{\Phi_g^{!\sim R\eta^{-1/\alpha}}} }(\xi)&=\bE_{\cdot|R}\left[e^{{-s\sum_{X_i\in\Phi_g^{!\sim R\eta^{-1/\alpha}}}p_gH\|X_i\|^{-\alpha}}}\right]\nnb\\
&\stackrel{(a)}{=}\bE_{\Phi_g|R}\left[\bE_{H|R}\left[e^{-s\sum_{X_i\in\Phi_g^{!\sim R\eta^{-1/\alpha}}}p_gH\|X_i\|^{-\alpha}}|\Phi_g\right]\right]\nnb\\
&\stackrel{(b)}{=}\bE_{\Phi_g|R}\prod_{X_i\in\Phi_g^{!\sim R\eta^{-1/\alpha}}}\frac{1}{1+p_g\xi \|X_i\|^{-\alpha}}\nnb\\
&\stackrel{(c)}{=}\bE_{U'|R}\prod_{(z_1,z_2)}^{ \mathbb{Z}^2}\frac{1}{1+p_g\xi||u'+s(z_1,z_2)||^{-\alpha}}\nnb\\
&=\int_{S_0\setminus B(0,R\eta^{-1/\alpha})}\prod_{(z_1,z_2)}^{\mathbb{Z}^2}\frac{1}{1+p_g\xi\|v+s(z_1,z_2)\|^{-\alpha}}\bP_{U
	'}(\diff v),
\end{align}
where (a) is obtained by conditioning on the locations of the grid, (b) is obtained by the Laplace transform of an exponential distribution with parameter one, and (c) follows from the fact that \emph{all} points of grid are measurable w.r.t the uniform random variable $ U'$ that takes a value uniformly in the set $ S_0'= S_0\setminus B(0,R\eta^{-1/\alpha}) $. 
\par On the other hand, we have 
\begin{align}
\mathcal{L}_{I_{\Phi_g\setminus \|U\|}}(\xi)&=\bE_{\cdot|U}\left[e^{{-s\sum_{X_i\in\Phi_g\setminus \{U\}}}p_gH/\|X_i\|^\alpha}\right]\nnb\\
&=\bE_{\cdot|U}\left[\bE\left[e^{-s\sum_{X_i\in\Phi_g\setminus \{u\}}p_gH\|X_i\|^{-\alpha}}\right]\right]\nnb\\
&=\prod_{(z_1,z_2)\neq (0,0)}^{\bZ^2}\bE_{\cdot|U}[e^{-sHp_g\|X_i\|^{-\alpha}}]\nnb\\
&\stackrel{}{=}\prod_{(z_1,z_2)\neq(0,0)}^{\mathbb{Z}^2}\frac{1}{1+p_g\xi\|U+s(z_1,z_2)\|^{-\alpha}}.\label{66}
\end{align}
Note that given that the typical user is associated with $ \|U\| $, the nearest BS of the grid does not comprise the interference. Since $ \Phi_g $ is \emph{measurable} w.r.t the associated BS at  $ U$, we have \eqref{66}

\section{Proof of Theorem \ref{Theorem2}}\label{A:1}
The expressions \eqref{eq:P5-0} and \eqref{eq:L4-1} evaluated at $ \xi=T\|u\|^\alpha p_g^{-1} $ with \begin{equation}
\bP(R_g>r\eta^{-1/\alpha})=\frac{\nu_2(S_0\setminus B(0,r\eta^{-1/\alpha}))}{\nu_2(S_0)} =\frac{\nu_2(S')}{\nu_2(S_0)},
\end{equation} gives $ \bP(\SIR>T,\cA_g). $
Similarly,	expressions \eqref{eq:P5-1} and \eqref{eq:L4-0} evaluated at $ \xi=Tr^\alpha p_p^{-1} $ with  \begin{equation}
\bP(R_p>\|u\|\eta^{1/\alpha})=e^{-\pi\lambda_p \|u\|^2\eta^{2/\alpha}} 
\end{equation} give  $ \bP(\SIR>T,\cA_p) $.  Therefore, the coverage probability is given by 
	\begin{align}
		&p_\textrm{cov}=\int_{S_0}{e^{-\pi\lambda_p \|u\|^2\eta^{2/\alpha}}}e^{\left.-2\pi\lambda_p\int_{\|u\|\eta^{1/\alpha}}^\infty\frac{T\eta\|u\|^\alpha v^{1-\alpha}}{1+T\eta \|u\|^\alpha v^{-\alpha}}\diff v\right.}\left(\prod_{{(z_1,z_2)\neq (0,0)}}^{\bZ^2}\frac{1}{1+\frac{T\|u\|^\alpha}{||u+s(z_1,z_2)||^\alpha}}\right)\bP_U{(\diff u)}\nnb\\
		&+\int_{0}^\infty {\frac{\nu_2(S_0')}{\nu_2(S_0)}}e^{\left.-2\pi\lambda_p\int_{r}^\infty\frac{Tr^\alpha v^{1-\alpha}}{1+Tr^\alpha v^{-\alpha}}\diff v\right.}\left(\int_{S'}\prod_{(z_1,z_2)}^{\bZ^2}\frac{}{1+\frac{\eta^{-1}Tr^\alpha}{||v+s(z_1,z_2)||^\alpha}}\bP_{U'}(\diff v)\right)f_{R}(r)\diff r\label{eq:theorem2-0}
	\end{align}
\par Furthermore, by change of variable, the first term is given by
	\begin{align}
	&\int_{S_0}{e^{-\pi\lambda_p \|u\|^2\eta^{2/\alpha}}}\prod_{{(z_1,z_2)\in \bZ^2\setminus (0,0)}}\frac{1}{1+\frac{T\|u\|^\alpha}{||u+s(z_1,z_2)||^\alpha}}e^{\left.-2\pi\lambda_p\int_{\|u\|\eta^{1/\alpha}}^\infty\frac{T\eta||u||^\alpha v^{1-\alpha}}{1+T\eta \|u\|^\alpha v^{-\alpha}}\diff v\right.}\bP_U{(\diff u)}\nnb\\
	&\stackrel{u=s\hat{u}}{=}\int_{\hat{S_0}}{e^{-\pi \rho_\lambda\|\hat{u}\|^2\eta^{2/\alpha}}}\prod_{{(z_1,z_2)\in \bZ^2\setminus (0,0)}}\frac{1}{1+\frac{T\|\hat{u}\|^\alpha}{\|\hat{u}+(z_1,z_2)\|^\alpha}}e^{\left.-2\pi\lambda_p\int_{s\|\hat{u}\|\eta^{1/\alpha}}^\infty\frac{s^\alpha T\eta\|\hat{u}\|^\alpha v^{1-\alpha}}{1+T\eta s^{\alpha}\|\hat{u}\|^\alpha v^{-\alpha}}\diff v\right.}\bP_{\hat{U}}(\diff \hat{u})\nnb\\
	&\stackrel{v=sw}{=}\int_{\hat{S_0}}{e^{-\pi \rho_\lambda\|\hat{u}\|^2\eta^{2/\alpha}}}\prod_{{(z_1,z_2)\in \bZ^2\setminus (0,0)}}\frac{1}{1+\frac{T\|\hat{u}\|^\alpha}{\|\hat{u}+(z_1,z_2)\|^\alpha}}e^{\left.-2\pi\rho_\lambda\int_{\|\hat{u}\|\eta^{1/\alpha}}^\infty\frac{T\eta\|\hat{u}\|^\alpha w^{1-\alpha}}{1+T\eta s^{\alpha}\|\hat{u}\|^\alpha w^{-\alpha}}\diff w\right.}\bP_{\hat{U}}(\diff \hat{u}),\nnb
	\end{align}
	where $ \hat{U}=\textrm{Uniform}([-\frac{1}{2},\frac{1}{2}]\times[-\frac{1}{2},\frac{1}{2}]) $ and then $ \bP_{\hat{U}} (\diff u)=1\cdot \diff x \diff y $.  
	\par In the same vein, the second term is  given by
	\begin{align}
	&\int_{0}^\infty{\frac{\nu_2(S_0')}{\nu_2(S_0)}} e^{-2\pi\lambda_p\int_{r}^\infty\frac{Tr^\alpha v^{1-\alpha}}{1+Tr^\alpha v^{-\alpha}}\diff v}\left(\int_{S_0\setminus B(0,r\eta^{-1/\alpha}) }\prod_{(z_1,z_2)}^{\mathbb{Z}^2}\frac{1}{1+\frac{\eta^{-1}Tr^\alpha}{||u'+s(z_1,z_2)||^\alpha}}\bP_{U'}(\diff u')\right)f_{R}(r)\diff r\nnb\\
	&\stackrel{u'=s\hat{u}'}{=}\int_{0}^\infty {\frac{\nu_2(S_0')}{\nu_2(S_0)}}e^{\left.-2\pi\lambda_p\int_{{r}}^\infty\frac{ T{r}^\alpha v^{1-\alpha}}{1+T{r}^\alpha v^{-\alpha}}\diff v\right.}\left.\int_{\hat{S_0}\setminus B(0,\frac{r\eta^{-1/\alpha}}{s})}\prod_{(z_1,z_2)}^{\mathbb{Z}^2}\frac{s}{1+\frac{s^{-\alpha}\eta^{-1} T{r}^\alpha}{||\hat{u}'+(z_1,z_2)||^\alpha}}\bP_{\hat{U}'}(\diff \hat{u}')\right.f_{R_p}(r)\diff r\nnb\\
	&\stackrel{r=s\hat{r}}{=}\int_{0}^{\infty}{\frac{\nu_2(\hat{S_0'})}{\nu_2(\hat{S_0})}}2\pi\rho_\lambda \hat{r}e^{-\pi \rho_\lambda \hat{r}^2}e^{\left.-2\pi\lambda_p\int_{s\hat{r}}^\infty\frac{s^\alpha T\hat{r}^\alpha v^{1-\alpha}}{1+s^\alpha T\hat{r}^\alpha v^{-\alpha}}\diff v\right.}\!\!\!\!\!\!\!\!\!\int\limits_{\hat{S}_0\setminus B(0,\hat{r}\eta^{-1/\alpha})}\prod_{(z_1,z_2)}^{\mathbb{Z}^2} \frac{1}{1+\frac{\eta^{-1}T\hat{r}^\alpha}{\|\hat{u}+(z_1,z_2)\|^{\alpha}}}\bP_{\hat{U}'}(\diff \hat{u}')\nnb\\
	&\stackrel{v=sw}{=}\int_{0}^{\infty}{\frac{\nu_2(\hat{S_0'})}{\nu_2(\hat{S_0})}}2\pi\rho_\lambda \hat{r}e^{-\pi \rho_\lambda \hat{r}^2}e^{\left.-2\pi\rho_\lambda\int_{\hat{r}}^\infty\frac{T\hat{r}^\alpha w^{1-\alpha}}{1+T\hat{r}^\alpha w^{-\alpha}}\diff w\right.}\!\!\!\!\!\!\!\!\!\int\limits_{\hat{S}_0\setminus B(0,\hat{r}\eta^{-1/\alpha})}\!\!\prod_{(z_1,z_2)}^ {\mathbb{Z}^2}\frac{1}{1+\frac{\eta^{-1}T\hat{r}^\alpha}{\|\hat{u}+(z_1,z_2)\|^{\alpha}}}\bP_{\hat{U}'}(\diff \hat{u}')\diff \hat{r}\nnb,
	\end{align}   
	where $ \hat{U}'=\textrm{Uniform}( [-\frac{1}{2},\frac{1}{2}]\times[-\frac{1}{2},\frac{1}{2}]\setminus B(0,\hat{r}\eta^{-1/\alpha}))$.
	Consequently, the coverage probability of the typical user is given by a function of $ \rho_\lambda $ and  $ \eta $. 
	
\section{Proof of Proposition \ref{lemma:bounds2}}\label{A:2}
	First, we develop upper and lower bounds for the Laplace transform of the interference given in \eqref{eq:49}. Note that the interference is comprised of signals from both grid and PPP. The Laplace transform of interference created by the PPP is well-known and does not require a bound. We need to find the upper and lower bounds for the interference created by the grid in \eqref{eq:49}. We have
	\begin{align}
	\cL_{I_{\Phi_g\setminus \{U\}}}(\xi)&=e^{\sum_{(z_1,z_2)\in \mathbb{Z}^2\cap W \setminus (0,0)}\log\frac{1}{1+\xi p_g\|U+s(z_1,z_2)\|^{-\alpha}}}\nnb\\
	&\geq e^{-\sum_{(z_1,z_2)\in \mathbb{Z}^2\cap W \setminus (0,0)}\xi p_g\|U+s(z_1,z_2)\|^{-\alpha}}\label{eq:L8-2},
	\end{align}
	where we use $ \log\left(\frac{1}{1+x}\right) \geq -x  $ for $ x\in (0,\infty)$. 
	\par In order to find an upper bound for \eqref{eq:49}, we define a measurable function $ \varphi (\cdot) $ such that 
	\begin{equation}
	\varphi(U,\xi)= \prod_{(z_1,z_2)\in \mathbb{Z}^2}\frac{1}{1+{\xi p_g}  ||U+s(z_1,z_2)||^{-\alpha}}\nnb.
	\end{equation} 
	Since the element  $ {1}/({1+{\xi p_g}  \|U+s(z_1,z_2)\|^{-\alpha}}) $  takes a value between zero and one, we have
	\begin{align}
	\cL_{I_{\Phi_g\setminus \{U\}}}(\xi)&=\varphi(U,\xi)\leq \prod_{(z_1,z_2)\in\bZ^2\cap W}\frac{1}{1+\xi p_g\|U\|^{-\alpha}}\label{eq:L8-3},
	\end{align} 
	where the inequality follows by truncating the function $ \varphi(U,\xi) $ by finite terms. We denote the truncation window by $ W $. 
	Applying \eqref{eq:L8-2} or \eqref{eq:L8-3} into \eqref{eq:49}, we obtain the lower and the upper bounds, respectively.

	\par Secondly, we develop lower and upper bounds for the Laplace transform of the interference created by the grid. In order to find the lower bound for \eqref{eq:50}, we have 
	\begin{align}
	\mathcal{L}_{I_{\Phi_g^{!\sim R\eta^{-1/\alpha}}} }\! (\xi)
	&=\bE_{\cdot|R}\left[e^{\sum\limits_{X_i\in\Phi_g^{!\sim R\eta^{-1/\alpha}}}\log\left(\frac{1}{1+\xi p_g \|X_i\|^{-\alpha}}\right)}\right]\nnb\\
	&=\bE_{\cdot|R}\left[e^{-\int_{\bR^2}\log\left({1+\xi p_g \|x\|^{-\alpha}}\right)\Phi_g^{!\sim R\eta^{-1/\alpha}}(\diff x)}\right]\nnb\!\!,
	\end{align}
	Moreover, 
	\begin{align}
	&\bE_{\cdot|R}\left[\exp\left(-\int_{\bR^2}\log\left({1+\frac{\xi p_g}{\|x\|^\alpha} }\right)\Phi_g^{!\sim R\eta^{-1/\alpha}}(\diff x)\right)\right]\nnb\\	&\geq\exp\left(-\bE_{\cdot|R}\left[\left. \int_{\bR^2}\log\left({1+\frac{\xi p_g}{ \|x\|^{\alpha}}}\right)\Phi_g^{!\sim R\eta^{-1/\alpha}}(\diff x)\right.\right]\right)\nnb\\
	&=\exp\left(-\int_{\bR^2}\log\left({1+\frac{\xi p_g}{\|x\|^\alpha}}\right)\bE[\Phi_g^{!\sim R\eta^{-1/\alpha}}(\diff x)]\right)\label{eq:campbell},
	\end{align}
	where we use Jensen's inequality and Campbells mean value formula \cite{baccelli2009stochastic}. We denote the conditional intensity measure of the grid by $ {\mathcal{M}}_{\Phi_g^{!\sim R\eta^{-1/\alpha}}}(\diff x) $ and it is 	
	\begin{align}
	\mathcal{M}_{\Phi_g^{!\sim R\eta^{-1/\alpha}}}(\diff x)&=\frac{\mathbbm{1}\{x\notin \bigcup_{X_i\in s\cdot \bZ^2}B(X_i,R\eta^{-1/\alpha})\}}{\nu_2(S_0 \setminus B(R\eta^{-1/\alpha}))}\nnb \diff x\\ &\leq \frac{\mathbbm{1}\{x\in\bR^2\setminus B(0,R\eta^{-1/\alpha})\}}{\nu_2(S_0 \setminus B(0,R\eta^{-1/\alpha}))}\diff x\label{eq:campbell2}.
	\end{align}
	In other words, {the intensity measure is upper bounded by the same uniform measure in $ S_0\setminus B(0,r\eta^{-1/\alpha}) $ that is nonzero for all the grid points except the origin.} By inserting \eqref{eq:campbell2} into \eqref{eq:campbell}, we have 
	\begin{equation}
	\cL_{I_{\Phi_g^{!\sim r\eta^{-1/\alpha}}}}({\xi})\geq e^{-\frac{2\pi\int_{R\eta^{-1/\alpha}}^{\infty}u\log\left({1+\frac{\xi p_g}{u^\alpha}}\right)\diff u}{\nu_2(S_0\setminus B(0,R\eta^{-1/\alpha}))}}\label{eq:L5-9}.
	\end{equation}
	\par In order to find the upper bound for \eqref{eq:50}, we use the measurable function $ \varphi(\cdot) $. Then,  
	\begin{align}
	\varphi(U',\xi)= &\prod_{(z_1,z_2)\in \mathbb{Z}^2}\frac{1}{1+{\xi p_g}  ||U'+s(z_1,z_2)||^{-\alpha}}\\&\leq \prod_{(z_1,z_2)\in \bZ^2\cap W}\frac{1}{1+\xi p_g \|U'+s(z_1,z_2)\|^{-\alpha}}\nnb,
	\end{align}by considering the window $ W $. Using    $\varphi(U')\leq\varphi_{\textrm{trun}}(U') \implies \bE[\varphi(U)] \leq \bE[\varphi_{\textrm{trun}}(U')]$, we have 
	\begin{align}
	\cL_{I_{\Phi_g^{!\sim r\eta^{-1/\alpha}}}}({\xi})&\leq\int_{S_0 \setminus B(0,R\eta^{-1/\alpha})}\prod_{(z_1,z_2)\in\bZ^2\cap W}\frac{1}{1+\xi p_g \|u+s(z_1,z_2)\|^{-\alpha}}\bP_{U'}(\diff u)\label{eq:81}.
	\end{align}
	 Applying the expression given in \eqref{eq:L5-9} or \eqref{eq:81} to  \eqref{eq:50}, we obtain the lower and upper bounds.

\bibliographystyle{IEEEtran}
\bibliography{ref}

\end{document}